\documentclass{aastex} 
\usepackage{graphicx,pdflscape} 
\pdfoutput=1

\slugcomment{Reference: Bianchi, L. 2024, ApJS, in press, DOI: 10.3847/1538-4365/ad6e7c}

\shorttitle{GUVcat\_AISxSDSS\_HS. Galactic Hot Stars in the GALEX UV surveys}
\shortauthors{Bianchi Luciana}

\newcommand{\fbina}{\mbox{b$_f$}}

\newcommand{\Lya}{\mbox{Ly$_{\alpha}$}}

\newcommand{\Teff}{\mbox{$T_{\rm eff}$}~}
\newcommand{\teff}{\mbox{$T_{\rm eff}$}}
\newcommand{\Lbol}{\mbox{$L_{\rm bol}$}~}
\newcommand{\lbol}{\mbox{$L_{\rm bol}$}}

\newcommand{\Rstar}{\mbox{$R_{\ast}$~}}

\newcommand{\Rsun}{\mbox{$R_{\odot}$}}

\newcommand{\Msun}{\mbox{$M_{\odot}$}}

\newcommand{\logg}{log~{\it g}}

\newcommand{\as}{\mbox{$^{\prime\prime}$~}}
\newcommand{\am}{\mbox{$^{\prime}$~}}

\newcommand{\rv}{$R_V$~}

\newcommand{\beqa}{\begin{eqnarray}} 
\newcommand{\eeqa}{\end{eqnarray}}

\newcommand{\ebv}{\mbox{$E_{B\!-\!V}$}}
\newcommand{\Ebv}{\mbox{$E_{B\!-\!V}$}~}
\newcommand{\efn}{\mbox{$E_{FUV\!-\!NUV}$}}

\newcommand{\Xtwo}{\mbox{$\chi^2$}~}

\begin{document}

\title{Hot Stars in the GALEX Ultraviolet Sky Surveys (GUVcat\_AISxSDSS\_HS) and the Binary Fraction of Hot Evolved Stars}

\author{Luciana Bianchi\altaffilmark{1}} 
\altaffiltext{1}{Dept. of Physics \& Astronomy, The Johns Hopkins University, 3400 N. Charles St.,  Baltimore, MD 21218, USA; 
http://dolomiti.pha.jhu.edu}
\email{bianchi@jhu.edu}

\begin{abstract}
We present a catalog of 71,364  point-like UV sources with SDSS photometry and GALEX FUV-NUV$\leq$0.1mag. 
The limit corresponds to stellar \Teff$\gtrsim$15,000$-$20,000K, slightly depending on gravity but nearly reddening-independent for Milky-Way-type dust. 
 Most  sources are hot white-dwarfs (WDs) and sub-dwarfs (SDs). Comparing the SED (GALEX~FUV, NUV, SDSS {\it u,g,r,i,z}) of 
 35,294 sources having good photometry  with colors of stellar models and known objects, 
 we identify 12,404$\pm$$^{1871}_{1267}$
  binary  hot-compact stars with a cooler, less-evolved companion (with a possible 8\%$-$15\% contamination by low-redshift QSOs), and  22,848$\pm$$^{1267}_{3853}$  single-star candidates. 
Single-star counts are an upper limit because pairs of similar stars have single-star-like SED, 
and  hot-WDs with main-sequence companions  of certain types 
(depending on WD's radius) are missed or counted as single 
in the available wavelength range and selection. 
The catalog offers unique leverage for identifying hot WDs,  elusive at longer wavelengths when a cooler, larger companion dominates optical-IR fluxes: 51\% of the binary- and 20\% of the single-star candidates are previously un-known objects.  
Gaia DR3 provides a parallax with error$\leq$20\%  for 34\%  of the binaries- and 45\% of single-star candidates, allowing 
\teff, \ebv, radius and \Lbol to be derived from SED analysis.  The binary-candidate sample
usefully expands the overall current binary-WD census to subpopulations elusive to Gaia and to other searches. 
 The binary fraction among this specific sample of hot-compact objects, albeit with the mentioned biases,
 \fbina $\gtrsim$46\%, compared with that of their progenitors
($>$80\%$-$50\% for mass range  8$-$1\Msun, \citet{moe2019}), implies a lower merging rate than found
 for massive stars by \citet{Sanaetal2017}.

\end{abstract}


\section{Introduction. Leveraging the GALEX UV Surveys to Identify Optically-Elusive Hot Stellar Sources.} 
\label{s_intro}

Hot stars are difficult to identify and to precisely characterize at optical wavelengths because their optical colors are saturated at  the \Teff of the earlier spectral types. Hot white dwarfs  (WDs) are  especially elusive at all wavelengths except the UV, due to their small radii and low optical luminosity. These hottest stars  stand out,  and their physical parameters are  easily characterized, if UV measurements are available, in particular at far-UV (FUV) wavelengths and shortwards.  UV colors not only are more sensitive to the hottest \teff's than  optical-IR colors are, but -  combined with data at longer wavelengths - they make it possible to remove the known degeneracy between \Teff and \Ebv for hot stars from  SED analysis, and to derive an unbiased \Teff value concurrently with extinction (e.g., \citet{bia14m31} - their Figure 1, \citet{bia11a,bia11b,bia18b}). 

The Galaxy Evolution Explorer (GALEX, \citet{martin05,morrissey07,bia09}) has surveyed the sky at ultraviolet (UV) wavelengths 
 for almost a decade. 
GALEX has imaged most of the sky 
in two Ultraviolet bands, FUV  ($\lambda$$_{eff}$ $\sim$ 1528\AA, 1344-1786\AA) and  NUV ($\lambda$$_{eff}$ $\sim$ 2310\AA,  1771-2831\AA)
simultaneously, 
until the FUV detector stopped working in May 2009, and observations continued with the NUV
detector only. With a field of view of $\approx$1.2$^{\circ}$  diameter 
and a spatial resolution of  $\approx$ 4.2\as (FUV) and 5.3\as (NUV) \citep{morrissey07},
 GALEX performed nested surveys with different area coverage and depth, see e.g., 
\citet{bia09,bia11a,bia14uvsky,bia17guvcat}; these works show also maps of the sky coverage of the surveys. 
The resulting photometric database contains about 600~million 
 UV measurements of almost 300~million UV sources,
 and $>$120,000 UV spectra \citep{bia18b}.  The GALEX database is a  unique 
resource for studies of hot
  stars and other classes of objects, including extra-Galactic objects such as star-forming galaxies and
  redshift$\lesssim$2 QSOs. 
Given that there was essentially no precursor all-sky or wide-area coverage UV  imaging survey \citep{bia2016a,bia2016b} and that a better one is still a few  years away at best, the GALEX UV sky surveys remain for now a unique resource for statistical studies of hot stellar sources. 

On average,  only about 10\% of the NUV sources are detected also in FUV, 
owing to the stellar IFM being  skewed towards cooler (less massive) stars, and UV-bright galaxies being more rare than red ones. For detailed maps of source content of the  sky as a function of UV magnitudes and colors, see  \citet{bia14uvsky} and \citet{bia17guvcat}.

The GALEX database\footnote{The GALEX data, including integrated images and source catalogs at single- and multi-visit (coadd) depths, are archived in the Mikulski Archive for Space Telescopes (MAST) 
 archive at the Space Telescope Science Institute (STScI), that also hosts the afore-mentioned $GUVcat$, $GUVmatch$, and other extracted catalogs in the Casjobs context ``GALEX\_catalogs''}
 contains  repeated measurements of some sources, when  fields have been re-observed or partly overlap.
Science-enhanced catalogs of unique UV sources, i.e. removing duplicate measurements, have also been produced
(\citet{bia11a,bia11b,bia14uvsky} 
  ($BSCcat$), \citet{bia17guvcat} ($GUVcat$, revised in 2020).
These catalogs support a variety of statistical studies and especially the matching
with other databases, which requires a unique source list without duplicate entries. To create a list of unique UV sources, in $GUVcat\_AIS$ \citet{bia17guvcat} associated  all existing 
measurements of the same source that may appear  with distinct GALEX IDs in the database for the All-sky Imaging Survey (AIS), the survey with the largest area coverage (tags PRIMGID and GROUPGID). When constructing $GUVcat$, \citet{bia17guvcat} also identified and corrected a number of bad ``co-adds'' in the GALEX database, that had resulted in wrong exposure
time and $FOV\_RADIUS$ values in tables such as $photoobjall$ in the GALEX database. 
$GUVcat$ also includes tags to facilitate science analysis, notably  INLARGEOBJ and
LARGOBJSIZE, that flag sources in the footprint of extended objects larger than 1\arcmin, 
such as large galaxies or stellar clusters, where crowding and
underlying unresolved galaxy light may compromise source photometry. \citet{bia17guvcat}
 noted that 1\am is a very conservative limit, for the purpose of eliminating crowded regions, but a user can choose to worry only about larger objects by using a combination of these two tags.\footnote{ \citet{bia17guvcat} also 
provided a list of the extended objects used to create these flags, and  finding charts for all of the extended objects ($>$1\am) in the footprint of GUVcat\_AIS. These can be found in the $GUVcat$ tools  web site 
\url{http://dolomiti.pha.jhu.edu/uvsky/\#GUVcat}.}

In this work we present a science-enhanced catalog of (mostly) hot stellar sources (Section \ref{s_sample}), with corollary data from SDSS, Gaia (Section \ref{s_gaia}) and Simbad (Section \ref{s_simbad}). In Section \ref{s_analysis} we perform bulk analysis of the sources, after culling the sample for quality (Section \ref{s_culling}). By comparing the sources seven-band SED (GALEX FUV , NUV, SDSS {\it u, g, r, i, z}) with model colors and with a subsample of known objects with classification, we identify binaries consisting of a hot compact object and a cooler, less-evoled star (Section \ref{s_binaccd}) and characterize some examples (Section \ref{s_sedfit}). The results are discussed in Section \ref{s_binaf} and summarized, along with the  features of the released catalogs of relevance for future users, in Section \ref{s_sum}.  Appendix~A contains details about the match of the UV sources  with their individual GALEX observations, and Appendix~B describes the released catalogs and where to find them.

\section{Selection of Hot Stellar Sources in the  UV Sky Surveys and Corollary Data.}
\label{s_phot}

\begin{figure}[h]
\includegraphics[angle=0,width=13.cm]{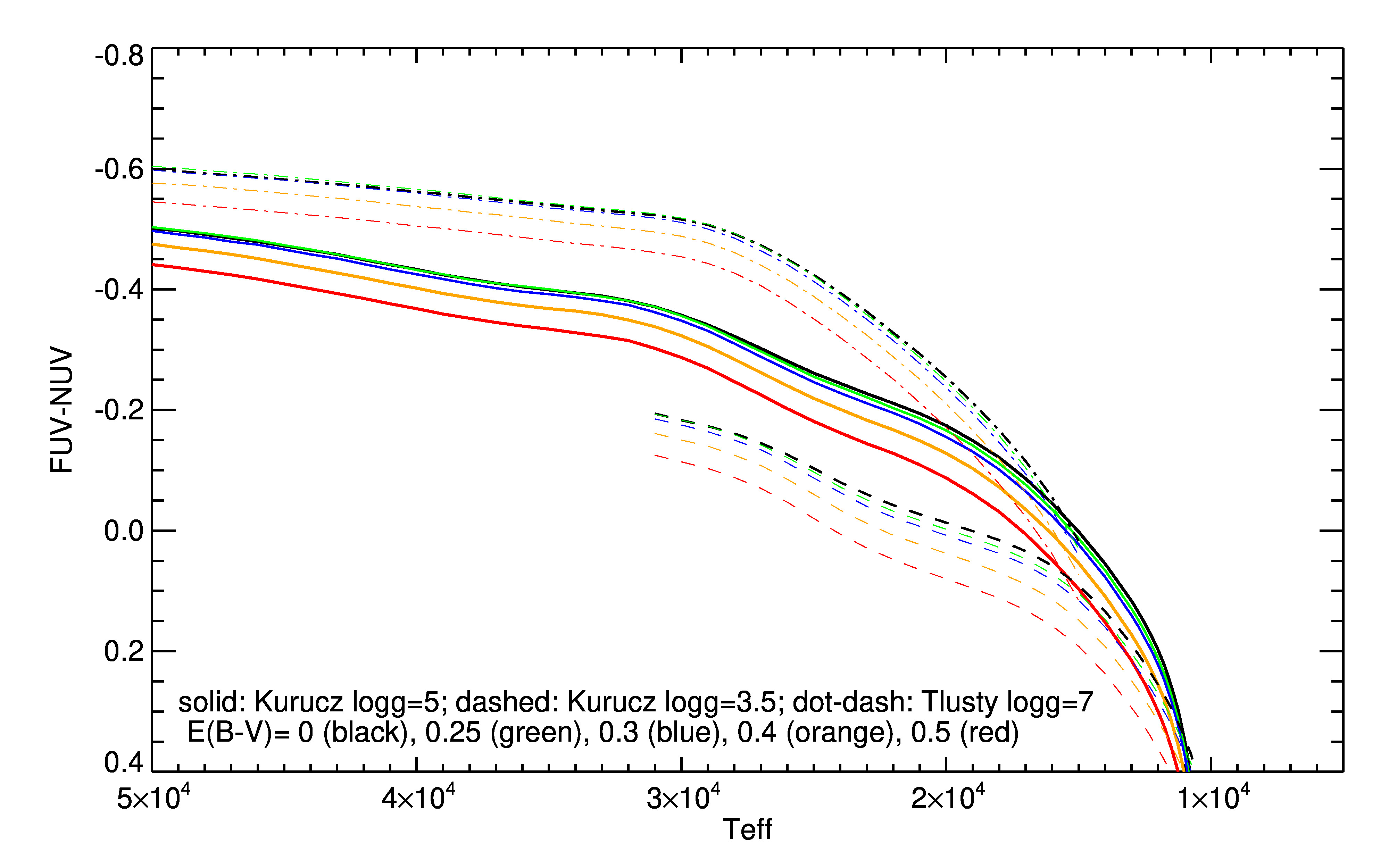}
\caption{Broad-band GALEX color FUV-NUV as a function of \Teff from stellar models: Kurucz models with solar metallicity, with \logg = 5.0 and 3.5, and Tlusty pure-H models with \logg = 7.0. The effect of reddening is negligible for small \Ebv values (for \ebv=0 and =0.25mag, shown in black and green, the lines are almost indistinguishable) and becomes appreciable at values higher than \ebv=0.40mag. Reddening is applied for a Milky-Way-type extinction curve with \rv=3.1. The reddening effect is larger for UV-steep extinction curves such as those found in the Magellanic Clouds (see e.g., Table~1 of \citet{bia17guvcat} ).    
 \label{f_fuvnuv} }
\end{figure}

\subsection{The Sample. Selection of Hot Stars from the GALEX  FUV-NUV Color} 
\label{s_sample}

The GALEX broad-band FUV-NUV is essentially a reddening-free color, for typical Milky-Way type extinction with \rv=3.1 (\citet{bia17guvcat} - their Table 1), because the very wide  NUV passband includes the 220nm extinction feature, which compensates the steeper extinction at FUV wavelengths by small dust grains.   Therefore, the GALEX color alone yields a robust selection of hot stellar sources, and a good indication of stellar \teff, nearly  independent of reddening, as illustrated in Figure \ref{f_fuvnuv}. 

However, to also characterize the type of the  hot stellar objects, and especially to identify those in binary systems with a cooler companion, we use as a starting point  the $GUVcat\_AIS$ catalog matched to the SDSS data release 14, $GUVmatch\_AISxSDSSdr14$ \citep{guvmatch}.   The match of the whole $GUVcat\_AIS$ ($\approx$83 million UV sources) with SDSS data release 14 (DR14) yielded 23,310,532 counterparts to 22,207,563 unique $GUVcat\_AIS$ sources, 10,167,460 of which are point-like \footnote{given the higher spatial resolution of SDSS, $\approx$1\as, compared to the 4.2/5.3\as (FUV/NUV) of GALEX, we use the SDSS definition of STAR/GALAXY in the SDSS photometric database, where ``STAR'' designates point-like sources, i.e. mostly stars and QSOs \citep{bia11a} and ``GALAXY'' generically refers to extended sources}, over a total overlap area of about 11,100 square degrees (area calculated with $AREAcat$, \citet{areacat}; sky-coverage maps are shown in \citet{bia14uvsky}). SDSS adds five optical magnitudes, {\it  u, g, r, i} and {\it z}, to the UV photometry.   \citet{guvmatch} had also estimated the statistical incidence of spurious matches, i.e.  positional coincidences of non physically associated sources. They matched the same SDSS database to a ``fake GUVcat'', i.e. a clone of the $GUVcat$ catalog where the position of the sources was offset by 5\arcmin, so to create a catalog of non-existing sources with the same magnitude and color distributions as the real UV-source catalog. The fraction of spurious matches turned out to be negligible, less than half percent in our pointlike sample; in more detail, as shown in their Figure 1 where the number of real and fake matches is compared, the fraction of spurious matches increases with match separation, as expected, because when including sources within a progressively increasing radius, the area increases with R$^2$, and the probability of finding random sources is higher in a larger area. On the contrary, the number of real matches is higher at short separations. This comparison was also used by \citet{guvmatch} to optimize the choice of match radius, as the best compromise to minimize both inclusion of accidental matches and loss of real matches, for an essentially complete matched catalog. 

We initially selected matched sources from \citet{guvmatch}'s $GUVmatch\_AISxSDSSdr14$ with FUV-NUV~$\leq$0.1mag and photometric error $\leq$0.3~mag in both GALEX filters, regardless of SDSS photometric errors or quality. The FUV-NUV~$\leq$0.1mag color cut corresponds to \Teff hotter than $\sim$15,000-20,000K, slightly depending on gravity and stellar type 
 but nearly independent of reddening unless the extinction is extremely high, as shown in  Figure \ref{f_fuvnuv};  \efn/\ebv=0.11 for Milky Way dust with \rv=3.1, see Table~1 of \citet{bia17guvcat}.   
The $GUVmatch\_AISxSDSSdr14$  catalog by \citet{guvmatch} includes all SDSS matches to a GALEX source within 3\as. We use the flag DISTANCERANK  to extract a unique source list, selecting the GALEX sources that have only one SDSS match (DISTANCERANK=0)  
and, for those with multiple SDSS matches, retaining the closest  match (DISTANCERANK=1).
These tags are propagated in the resulting catalog, because for UV sources that have additional SDSS matches within the match radius, i.e. that are resolved into multiple optical sources within 3\as,  the UV flux might  be composite of the optically-resolved matches, and the UV-optical color of the closest match may  be biased. Therefore, sources that have additional matches (DISTANCERANK=1) 
must be treated with caution. For sources of interest, the additional matches can be examined in the $GUVmatch\_AISxSDSSdr14$ catalog by \citet{guvmatch}. 
The FUV-NUV$\leq$0.1mag selection yields 278,433 sources (we refer to the initial selection as $GUVcat\_AISxSDSS\_HSall$). We eliminate 36 sources with FUV\_ARTIFACT=32 and 22 sources with  NUV\_ARTIFACT=32  (rim artifact, \citet{bia17guvcat}), resulting in a sample 278,375 GALEXxSDSS matched sources before further culling, which is discussed later.

FUV\_MAGs range from 11.72 to 22.52~ABmag in the sample, and NUV\_MAGs are  between 12.37 and 22.80~ABmag; therefore, the sample includes some sources in the non-linear or saturated count-rate regime. These are not eliminated upfront, because they might be interesting sources for follow-up with other instruments.  Non-linearity roll-off at 10\% level sets in at 13.73~ABmag for FUV and 13.85~ABmag for NUV, see \citet{morrissey07} and in particular their Figure 8.

As mentioned in Section \ref{s_intro}, $GUVcat\_AIS$'s tag INLARGEOBJ  flags sources that are in the footprint of extended objects such as galaxies or stellar clusters  larger than 1\am.
For a foreground stellar source in the line of sight of a bright galaxy disk or in a crowded cluster,
  the source identification and photometry can be extremely problematic, as  illustrated in Figure 5 of  \citet{bia17guvcat}. The initial $GUVcat\_AISxSDSS\_HSall$ 
sample includes 915 sources with tag INLARGEOBJ not 'N'; 
 these are in 133 distinct extended objects, of sizes ranging from 301\am to 1\am (see \citet{bia17guvcat} for definition of sizes)\footnote{listed in order of decreasing size:
  {\tiny  OC:MELOTTE111 OC:NGC2632 GA:NGC0224 SC:PLATAIS2 OC:NGC2682 GA:NGC0598 OC:NGC2548 GA:UGC10822 GC:NGC5272 GA:PGC088608 GC:NGC6341 SC:NGC7772 OC:MWSC4288 GA:NGC5457 OC:MWSC4301 SC:DOL-DZIM6
OC:ALESSI62 GA:NGC2403 OC:MWSC5154 SC:COLLINDER21 GA:IC1613 GC:NGC7089 GA:NGC4565 GA:NGC4244 OC:MWSC5033 OC:MWSC5901 OC:MWSC4602 OC:NGC2420 OC:ALESSI10 OC:MWSC4572 OC:MWSC4383 OC:FSR1064
OC:MWSC5804 OC:MWSC5828 OC:ASCC41 OC:MWSC5016 OC:FSR1102 GA:NGC4472 OC:FSR0080 GA:NGC0628 GA:NGC5033 SC:MWSC5044 OC:MWSC5018 SC:FSR0876 OC:FSR0765 OC:MWSC5735 SC:MWSC5038 OC:MWSC5076
SC:NGC6481 GA:UGC07698 OC:FSR0161 GA:NGC3945 GC:KOPOSOV2 GA:UGC02275 GA:NGC4579 GA:UGC02302 GA:UGC05829 GA:UGC01176 GA:UGC09242 GA:NGC4651 GA:UGC01133 GA:UGC07608 GA:IC3687
GA:SDSSJ113637.41+112327.0 GA:NGC6255 GA:NGC3104 GA:NGC4151 GA:NGC1199 GA:NGC4340 GA:UGC08331 GA:NGC5866B GA:UGC08651 GA:UGC09537 GA:UGC01547 GA:IC2329 GA:UGC05340 GA:UGC04499
GA:UGC08683 GA:UGC04837 GA:UGC00501 GA:UGC03966 GA:NGC0807 GA:UGC07307 GA:UGC07719 GA:PGC045321 GA:UGC08572 GA:PGC032620 GA:UGC09828 GA:IC3105 GA:PGC007998 GA:UGC05907 GA:IC2421
GA:UGC00035 GA:UGC08426 GA:NGC7469 GA:UGC01519 GA:UGC00052 GA:IC0949 GA:PGC022928 GA:UGC12388 GA:PGC3130476 GA:NGC0317A GA:NGC7603 GA:NGC5790 GA:PGC071675 GA:UGC07906 GA:UGC01697
GA:UGC01318 GA:PGC071916 GA:2MASXJ11052812+2500595 GA:PGC057881 GA:UGC06086 GA:UGC05999 GA:UGC04249 GA:PGC037722 GA:IC2568 GA:SDSSJ161622.19+041407.7 GA:UGC10025 GA:UGC00370 GA:PGC053682
GA:PGC2806871 GA:UGC04940 GA:SDSSJ114553.33+635411.6 GA:UGC06980 GA:2MASXJ09520676+2638256 GA:UGC06027 GA:UGC09458 GA:UGC09698 GA:PGC061304 GA:PGC023515 GA:PGC135848 GA:PGC060039
GA:PGC023508; \\ }
sizes in arcmin:  {\tiny
301.2,217.2,177.8,158.4, 66.0, 62.1, 56.4, 36.8, 32.4, 30.0, 28.8, 26.4, 24.0, 24.0, 22.8, 20.4, 20.4, 20.0, 19.2, 19.2, 18.3, 16.8, 16.7, 16.2, 15.6, 15.6, 15.6, 15.0, 15.0, 15.0, 13.8, 13.2, 13.2, 12.6, 11.4, 10.8, 10.8, 10.2, 10.0,  9.9,  9.8,  9.6,  9.6,  9.6,  9.6,  9.0,  8.4,  8.4,  7.8,  6.3,  6.0,  5.5,  5.2,  5.0,  5.0,  4.7,  4.5,  4.4,  4.2,  3.9,  3.5,  3.4,  3.3,  3.2,  3.1,  3.0,  2.9,  2.8,  2.8,  2.6,  2.3,  2.3,  2.2,  2.0,  2.0,  2.0,  1.9,  1.9,  1.9,  1.8,  1.7,  1.7,  1.7,  1.6,  1.6,  1.6,  1.6,  1.6,  1.5,  1.5,  1.4,  1.4,  1.4,  1.4,  1.4,  1.3,  1.3,  1.3,  1.3,  1.3,  1.3,  1.2,  1.2,  1.2,  1.2,  1.2,  1.2,  1.2,  1.1,  1.1,  1.1,  1.1,  1.1,  1.1,  1.1,  1.1,  1.1,  1.1,  1.1,  1.1,  1.1,  1.1,  1.1,  1.0,  1.0,  1.0,  1.0,  1.0,  1.0,  1.0,  1.0,  1.0,  1.0} }.  
 Several of the extended objects that contain sample sources are stellar clusters (INLARGEOBJ=OC:--- or GC:--- or SC:--- ;  see column 94 in Table 
 \ref{t_tagspoint}); we do not  discard the hot sources in clusters from the initial catalog, although the standard pipeline photometry is sometimes unreliable (see later). Out of the 133 unique extended objects including  $GUVcat\_AISxSDSS\_HSall$ sources, 59/68/74/78 have sizes $\leq$ 2\am/3\am/4\am/5\am, and all of these are galaxies.  Some $GUVcat\_AISxSDSS\_HSall$ sources are also found in larger galaxies: GA:NGC0224, GA:NGC0598, 
GA:UGC10822, GA:PGC088608,  GA:NGC5457, 
GA:NGC2403, GA:IC1613,  GA:NGC4565, GA:NGC4244, GA:NGC4472,  GA:NGC0628, GA:NGC5033, 
GA:UGC07698,  GA:NGC3945, GA:UGC02275, GA:NGC4579.  
GALEX images  of all the extended objects  larger than 5\am, with $GUVcat$ sources overlaid,  can be viewed in the $GUVcat$ web site.\footnote{http://dolomiti.pha.jhu.edu/uvsky/uvsky\_GUVcat.html , see  e.g.,\\ 
 http://dolomiti.pha.jhu.edu/uvsky/GUVcat/GUVcat\_AIS/ExtendedObjects\_GA\_page1.html }   
 Sources in large galaxies are likely  integrated fluxes of UV-bright regions (see Figure 5 of \citet{bia17guvcat}); therefore, they are  meaningless for our purpose and they will be discarded from the analysis of stellar hot sources. There might be bright stellar sources, such as novae, in nearby galaxies images, but these cases  would require performing custom photometry rather than using the automated pipeline extraction.

Of the 278,375 GALEX\_AISxSDSS matched sources with FUV-NUV$\leq$0.1mag, 71,364 are defined point-like in the SDSS database. SDSS {\it TYPE}='STAR' indicates `point-like' source morphology; therefore,  the source could be a star or a QSO. Of the remaining 207,011, ten have SDSS {\it type}='UNKNOWN' and the other 207,001 have {\it type}='GALAXY', indicating that they are  treated as extended sources by the SDSS pipeline.

In this work, we focus our analysis on the 71,364 sources deemed point-like  by the SDSS pipeline, because SDSS has $\approx$3$\times$ higher spatial resolution than GALEX images.  We refer to this subset as the master catalog  
(released as $GUVcat\_AISxSDSS\_HSpoint$), the starting point for the analysis.
 Of the 71,364 point-like sources, 360 are flagged to be in extended objects, mostly open clusters (INLARGEOBJ= OC:MWSC5735, OC:ALESSI10, OC:MWSC5828,  OC:FSR0161,   OC:FSR0765, OC:ALESSI62, OC:FSR0080, OC:NGC2548, OC:FSR1064, OC:NGC2420,  OC:MWSC4383, OC:MWSC4288, OC:NGC2632, OC:MWSC5154,   OC:MELOTTE111) 
or globular clusters (GC:NGC7089, GC:NGC6341, GC:NGC5272) or other stellar clusters (SC:NGC7772, SC:COLLINDER21, SC:PLATAIS2, SC:NGC6481,  SC:FSR0876, SC:DOL-DZIM6, SC:MWSC5038, SC:MWSC5044), but some sources measured as point-like by SDSS are also in the footprint of galaxies.\footnote{ {\tiny  GA:PGC007998, GA:PGC071675, GA:UGC01697, GA:UGC01133, GA:NGC062, GA:NGC0628, GA:UGC01519, GA:NGC0598 (most of the sources), GA:NGC0224, GA:PGC023515, GA:PGC3130476, GA:NGC2403,
    GA:UGC10822,   GA:IC2421,  GA:PGC088608,  GA:NGC5866B, GA:SDSSJ114553.33+635411.6, GA:NGC5790, GA:PGC135848, GA:IC2568, GA:NGC5457,  GA:PGC045321, GA:UGC08572, GA:NGC4472, GA:NGC4151, GA:IC0949, GA:UGC07719, GA:UGC06980, GA:UGC07698} }
Hot stars in stellar clusters are of course of great interest, but  require custom photometry, because the standard GALEX pipeline is not robust in crowded fields: see, e.g.,  \citet{demartino2420} and Figure 2 of \citet{bia14}. Therefore, all sources flagged in the footprint of extended objects  are kept in the $GUVcat\_AISxSDSS\_HSpoint$ catalog for completeness. Although most of these sources can probably be recovered by using the aperture photometry (included in the master catalog) or custom photometry, they are not used in the present analysis in favour of  a clean, homogeneous sample with consistent processing, in particular the pipeline defined ``best'' magnitude, recorded as FUV\_MAG and NUV\_MAG.   This initial point-like sample will be further culled for the intended analysis (Section \ref{s_culling}) but it is provided publicly in its entirety for possible other uses.

\subsection{Match with Gaia DR3} 
\label{s_gaia}

We matched the input catalog of 71,364 point-like sources $GUVcat\_AISxSDSS\_HSpoint$ with Gaia DR3 release \citep{gaiadr3} (before culling the sample for analysis, as will be discussed later in Section \ref{s_culling}). A 3\as match radius was used; this value was found by \citet{guvmatch} to be a good compromise to minimize both embarking spurious matches and missing real matches. For point sources, the difference between the GALEX position and the position of the Gaia match is usually much smaller than 3\as (\citet{guvmatch}, see their Figure 1).  Using a conservatively large  match radius, however,  is needed so that  possible close-by sources that may be unresolved by GALEX (making the UV$-$optical colors biased) are also recorded, as well as possible matches with high proper motion sources, at the risk of  embarking some spurious additional matches.  In the final match output, therefore, GALEX sources with multiple matches or with one match close to 3\as must be examined with caution.  

The epoch of Gaia (J2016) matched source and the date of the  GALEX observation (2003-2013) may differ by up to 13 years. Therefore,  a simple cross-match between coordinates from the two databases could include Gaia sources within 3\as of the GALEX source position that were slightly farther than 3\as at the epoch of the GALEX observation, or exclude sources that were closer than 3\as at the time of the GALEX observations, but beyond 3\as when measured by Gaia. Such cases are very rare, because they require a large proper motion. However, because we are dealing with a stellar sample, 
rather than simply cross-matching the input list with the Gaia DR3 J2016 positions we followed a more precise procedure, accounting for position shifts due to proper motion. We performed an initial match of the GALEX sources catalog to Gaia DR3 with a very large match radius (15\as) which is the maximum possible displacement in 13 years among Gaia sources, considering the highest proper motion values in the entire Gaia DR3.   We performed this initial cross-maching at the Casjobs interface at MAST.
 Then, for all potential Gaia matches within 15\as that have a significant proper motion measurement (a fraction of the matches), we used the proper motion (DR3 tags: PMRA, PMDEC) 
to ``regress'' the Gaia J2016 position to where it was at the epoch of the GALEX observation. The 15\as-radius cross-matching returned 95,400 
Gaia sources, including some multiple matches of the same  GALEX source; of these, 4,370 input sources have a Gaia match within 15\as but not within 3\as.  And 10,399 GALEX sources do not have a Gaia match even within 15\as.
 
 Regressing the Gaia position of each potential match to the time of the GALEX observation requires an additional step back from the GALEX extracted catalog. The sources were selected from the GALEX database that combines all GALEX observations of the same source (when multiple visits exist); therefore, the sources' RA,DEC values were measured on a combination (``coadd'') of all existing images of that source in most cases (see \citet{bia17guvcat} for more details). By consequence, the GALEX source position in the original database does not have a corresponding observation epoch (the misleading tag ``EPOCH''  in the Casjobs GALEX  database is actually  the equinox of the coordinate system, J2000). In order to obtain a reference position consistent with an observing date for the GALEX sources, we matched the sample  with all GALEX measurements from individual observations (GALEX ``visits''), by cross-matching our  $GUVcat\_AISxSDSS\_HSpoint$ catalog with  table $visitphotoobjall$ in the Casjobs GALEX database, that contains measurements from all GALEX data at individual-visit level (about 600 million source measurements). A 3\as match radius was used.  When multiple visits were found for a source, we selected the ``best'' visit, chosen as the observation with the longest exposure time, excluding those visits where the source was near the edge of the field, which suffers by some distortion.
 We then used the Ra,DEC,Epoch of the GALEX best visit for a refined Gaia matching.   The match of the input source list with the information at individual visit level is described in Appendix~A 
 and the full results are also made available, given that a number of sources have multiple observations, and these can be useful for different purposes, including a serendipitous  variability search between repeated observations, or for examining different images of specific sources of interest. Relevant parameters of the GALEX  ``best-visit'' are included in the master catalog (Table \ref{t_tagspoint}). For 25 input sources a visit-level match was not found within 3\as (see Section \ref{s_appendixA}); these are eventually excluded from the analysis by other culling criteria, as explained in Section \ref{s_culling}.

 Having chosen for each source a GALEX ``best visit'' from the individual observations, we recomputed   the spherical distance between  the best-visit's GALEX RA,DEC and all potential Gaia matches found within 15\as, with their DR3 position ``rewinded'' to the date of GALEX's ``best-visit'' (if a significant proper motion value exists).  The results from the Gaia match include, in the master catalog, two tags for the separation between the GALEX source position and the Gaia counterpart: SEPBACK\_ARCSEC (the separation from the GALEX best-visit position and the Gaia match accounting for proper motion correction, i.e. the Gaia source position at the epoch of the GALEX best visit) and DISTARCMIN (the separation as returned from the initial DR3 J2016 positional match). The ``rewinded'' Gaia position at the epoch of GALEX's best visit is also given, tags:  Gaia\_RAback,  Gaia\_DECback (while GAIA\_RA and GAIA\_DEC are the epoch=J2016 position from DR3). See Table \ref{t_tagspoint} for description of all tags.   When there is no significant proper motion, the two separations and the two Gaia positions coincide, and for the 25 input sources without a visit-level detection within 3\as, the proper-motion correction cannot be applied either. Of all Gaia sources within 15\as,   58,980 
have SEPBACK\_ARCSEC $\leq$3\as, i.e. are within a 3\as match radius using the position corrected for proper motion (58,269  
 of these have also DISTARCMIN*60. $\leq$3\as), and 58,538 
 have  DISTARCMIN*60. $\leq$3\as, regardless of SEPBACK\_ARCSEC, i.e. before applying proper motion corrections. Because the two cuts (i.e., accounting for proper motion or not) yield some sources that are not in common, we retain in the final Gaia matched catalog all matched sources that have separation $\leq$3\as either before or after accounting for proper motion, a total of 59,249 entries.  The matches can be trimmed to a smaller match radius using the tags DISTARCMIN and SEPBACK\_ARCSEC.     There are 427 sources that have a primary match within 3\as after registering the Gaia positions to the GALEX best-visit epoch using the proper motion,  but were beyond 3\as using the DR3 J2016  position. Their proper motion values range from 0.259288   to    965.74628~mas and the GALEX epochs between years 2003.4742    and   2011.7510 (year.decimal).  Also, 45 sources that are within 3\as using the Gaia J2016 position are farther than 3\as after applying the proper motion (PM range: 0.15602224    to   55.018314~mas, GALEX epoch: 2003.7570   to    2009.4041). 
  The 
sources that are within 3\as after correcting for proper motion but are not according to their J2016 position, or viceversa, include cases with high proper motion values or close to the 3\as match-radius cut off limit. For example, the highest discrepancy between the Gaia$-$GALEX  separation before and after proper-motion correction 
is for source GALEX\_ID=6373800939499817688; the Gaia match (Gaia~ID = 943770757800160384) has a proper motion of 965.746~mas; using the DR3 (epoch J2016) coordinates, the Gaia source    is 10.5335\as away from the GALEX input position, but after correcting for proper motion it was 1.6888\as away from the GALEX best-visit RA,DEC (epoch = 2006.9764, about 10 years earlier than the Gaia observations).  Its parallax of 58.5336~mas with a small error ($<$1\%), places the source at a distance of only 17.08~pc. 
 This case and other similar ones are retained because, in favour of completeness,  we did not limit proper motion corrections based on proper motion uncertainties. The uncertainties are propagated  in the master catalog, and can be examined to refine the sample according to specific needs and criteria. 
The above numbers 
do not include the 15,106 
GALEX input sources that do not have a Gaia match (10,399 without a match even within 15\as and 4,707 
 with a match within 15\as but beyond 3\as). 

Following \citet{guvmatch}, we built tags to track input sources with multiple Gaia matches within 3\as, as described in Table \ref{t_gaiatags}. These tags are also propagated in the master catalog (Table \ref{t_tagspoint}). In particular, MMRANK\_GAIA is defined as:
\begin{itemize}
\item
MMRANK\_GAIA = 0 if the Gaia match is the only one for the GALEX source (53,331 cases) 
\item
MMRANK\_GAIA = 1 if the  Gaia match is the closest of more than one match (2,927 cases) within the match radius 
\item
MMRANK\_GAIA $>$1 in order of increasing distance from the GALEX position, for additional Gaia matches (2,991 sources)); there are up to 4 matches for one source\footnote{these additional matches are only found in the full match result (Appendix~A);  in the master catalog only the primary match is propagated - to maintain a unique-source list - but this tag carries the information on whether additional matches exist} 
\item
MMRANK\_GAIA = -888 (as all other Gaia related tags) if there is no Gaia match within 3\as  (15,106 sources, 
about  21\% of the input master catalog)
\end{itemize}

When a GALEX input source has multiple Gaia matches within the match radius, all the matches are listed in the full output (Section \ref{s_cats}, i.e. Appendix~B), in subsequent rows where the GALEX input source is repeated. To retrieve the list of unique input sources, one can select MMRANK\_GAIA $\leq$1 from the full match output tables.  In the master catalog, only the closest match is retained, but the ``MMRANK\_GAIA'' and ``NMMRANK\_GAIA'' tags carry the information of whether multiple matches exist. 

Of the 59,249 input  
sources that have at least one Gaia match, i.e. the primary matches that are not ``null'' (null matches have GAIA\_ID = '-888' in the master catalog), 53,331 have only one match within 3\as, and only 2,927 have additional matches. Such figures imply that there is little ambiguity on the actual Gaia counterpart for most GALEX sources. 
Of these Gaia matches, 6,757 
have no parallax measurement, and of the 52,492 
matches with a parallax value in DR3, 46,443  have a parallax $>$0.~mas, while 6,049 have a negative parallax (a value resulting from DR3 pipeline, not null). Negative parallax values, when not due to large uncertainties, may indicate a failure of the solution parallax$-$proper-motion, which in our sample could be related to binarity, either owing to orbital motion not yet disentangled from proper motion in DR3, or to semi-resolved binaries causing a shift of the source centroid among the repeated Gaia observations.  Therefore, although a DR3 negative parallax prevents absolute estimates of the stellar parameters, possible failed-solution cases (negative parallax) are potentially interesting targets for the purpose of investigating very close binaries, augmenting the impressive results produced by Gaia on binaries that can be measured by Gaia (e.g., \citet{elbadry24gaia}).  As we will see, the present sample can especially stretch the census of binaries containing a hot compact object, that may be lacking in the Gaia binary-WDs population.

In sum, out of the  71,364 GALEX sources in the input catalog (before further culling for science analysis)
15,106 GALEX sources have no Gaia match; they are retained in the master catalog 
with values of Gaia tags SOURCE\_ID as well as all other Gaia-related  parameters set to -888.

If we now consider the 56,258  
  sources with either a unique Gaia match, or the closest match in case of multiple matches (i.e., we do not include additional matches beyond the closest match in the statistics),  
50,098 primary matches have a parallax value in Gaia DR3, 
but only 44,258 have parallax $>$0., and of these, 19,258
have a parallax error $\leq$20\%.  We will give particular emphasis in the analysis to the subsample for which a good direct distance estimate is available, so that absolute stellar parameters can be derived (Section \ref{s_analysis}), after further  sample cleaning in Section \ref{s_culling}. 

The Gaia DR3 matches to the GALEX sources have also been cross-linked with DR3 table $vari\_summary$;   the resulting full match output catalog (Section \ref{s_cats} (Appendix~B) ) includes all tags from both the main Gaia {\it source} table and the variability information when available, i.e. when variability has been detected by Gaia.  

The most relevant tags of the matched sources are included in the master catalog of point-like sources, as well as some distilled information on Gaia (serendipitous) variability from $vari\_summary$. The full matching results, with all tags from both Gaia $source$ and $vari\_summary$ are also available online as a separate file (Section \ref{s_cats}).

\subsection{Match with the SIMBAD database}
\label{s_simbad}

The input list of point-like hot GALEXxSDSS sources was also matched with the CDS Simbad database,  using the Vizier interface at \url{http://cdsxmatch.u-strasbg.fr}, specifically \url{http://cdsxmatch.u-strasbg.fr/\#tab=xmatch\&}.

 We used two values of match radius, 5\as and 10\as,  larger  than the value adequate for Gaia matching, because bright stellar objects with entries in Simbad from old catalogs may have rounded or less precise coordinates, and we aimed  at collecting a complete set of possible matches, to be culled later in the analysis if necessary.

The Simbad cross-match to $GUVcat\_AISxSDSS\_HSpoint$ returns 34,831  
and 35,665 
matches within match radius = 5\as and 10\as respectively; some of
these are multiple matches of the same source. That is, less than half
of the input point-like GALEXxSDSS hot sources  are known objects.
The full results from the Simbad matches are  given in separate tables
(see Section \ref{s_cats} i.e. Appendix~B).  Among all the Simbad matches within  match radius =5\as (=10\as), 
 2008 (2152) 
 have Simbad MAIN\_TYPE = Star,\footnote{
Simbad OBJECT\_TYPEs are described in \url{http://simbad.cds.unistra.fr/guide/otypes.htx}; see also \citet{simbadtypes} }
and 14,830 (14,842) 
have MAIN\_TYPE = WD*\_Candidate, to cite the more numerous classes.   These numbers highlight the power of UV data to identify hot WDs, by confirming the high percentage of WD candidates among the known objects on the one hand, and - on the other hand - by showing the limitation of  current known samples, and the capability of UV-data to significantly extend the census of such objects, which are elusive at other wavelengths. 
Only 34 (37) 
do not have Simbad tag OTHER\_TYPES set;  
 13,099 (13,137) 
 have a reported spectral type (Simbad tag SP\_TYPE, propagated in the master catalog as SIMBAD\_SP\_TYPE\_NEAREST), while about two thirds, 21,732 (22,528),
do not have a reported SP\_TYPE. These statistics refer to the full Simbad match results, including multiple matches. 
 These tags, and other information distilled from the Simbad match, are included in the master catalog (see Table \ref{t_tagspoint} for field description,  columns 223-231);  
if there are multiple matches to a GALEX source, in the master catalog the values are given for the nearest match, and additional matches can be found in the full output. The tags of the closest match propagated in the master catalog are extracted from the 5\as match-radius results; if there is no match within 5\as, the values are set to ``=='', even if  there is a match further away than 5\as and within 10\as, because the chance that such distant matches are spurious (positional coincidence) is high.
  The number of Simbad matches within 5\as and 10\as is also reported in the master catalog, in tags N\_SIMBAD\_MATCHES\_5AS and N\_SIMBAD\_MATCHES\_10AS respectively. Therefore, if there is no match within 5\as but there are matches within 10\as, or in case of multiple matches, these are not included in the master catalog, but the  tag 
 N\_SIMBAD\_MATCHES\_10AS $>$0  indicates that they exist and  can be retrieved in the complete Simbad cross-match output files for the sources of interest (see Section \ref{s_cats}).  
There are up to five (eleven) Simbad matches to one GALEX source within 5 (10)\as. 
 Of the 71,364 $GUVcat\_AISxSDSS\_HSpoint$ sources, 37,273 
 have no Simbad match within 5\as and 36,985 
do not have any match even within 10\as.  Such small difference  indicates that we can probably ignore matches beyond 5\as. In some cases,  multiple Simbad matches of a source could be the same object, that exists in the CDS database with different identifiers.
Of the known objects, 21,263 do not have a spectral type classification.

\section{Analysis}
\label{s_analysis} 

\subsection{Culling the Sample for Analysis} 
\label{s_culling}    

For the analysis that follows, we cull the 
$GUVcat\_AISxSDSS\_HSpoint$
catalog by eliminating sources that may have bad or inconsistent (between GALEX and SDSS) photometric measurements, resulting in a biased UV$-$optical SED, and other warnings.   First, we exclude all sources with flag INLARGEOBJ not equal 'N' (Section \ref{s_sample}), at the expense of eliminating interesting sources in stellar clusters; this criterion reduces the sample from  71,364 to 71,004 sources.  

\begin{figure}[!h]
\centering
\includegraphics[scale=0.11]{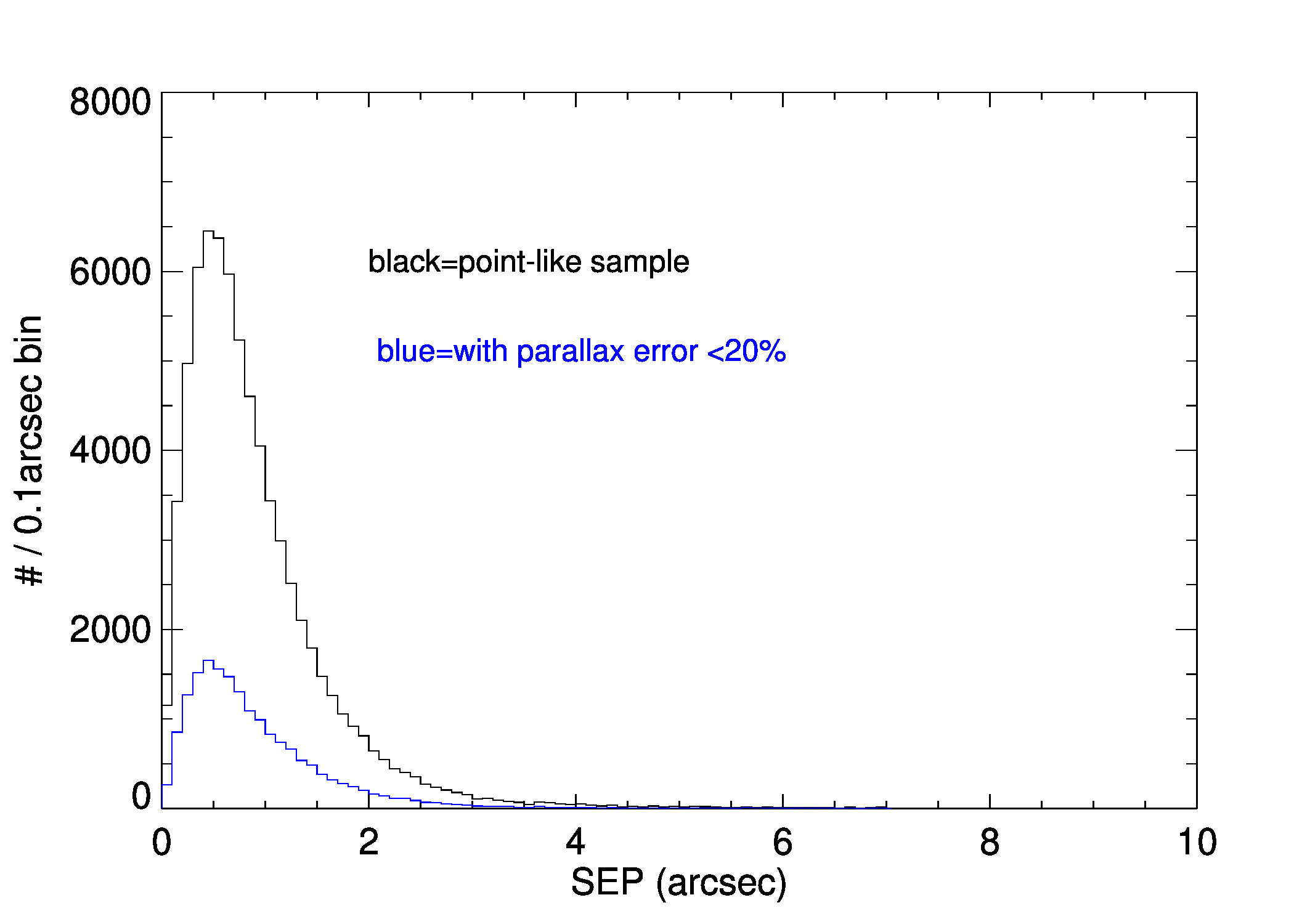}
\includegraphics[scale=0.11]{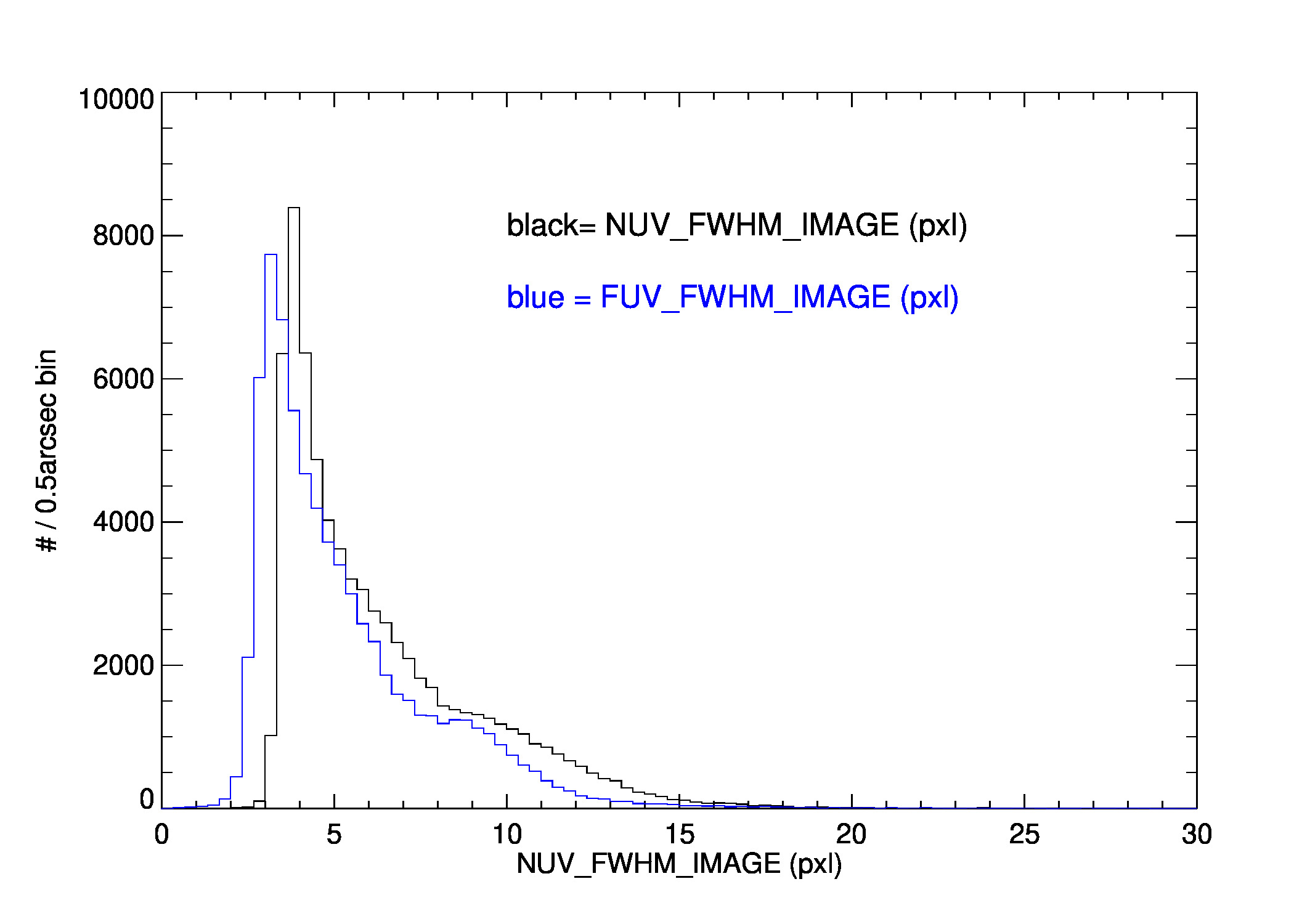}
\caption{
\small
{\it Left:} 
Distribution of the separation between FUV and NUV position of the extracted sources (``SEP'' tag from the GALEX pipeline); {\it Right:} Histogram of   [F/N]UV\_FWHM\_IMAGE (in pxl, 1pxl=1.5\as) measured by the GALEX pipeline.  There is no correlation between these two extraction parameters over the sample. 
\label{f_SEP_FWHM} }
\end{figure}

 Then, we examine the tag ``SEP'', the separation between the FUV and NUV source positions from the pipeline detection, that were merged into one GALEX source. For most GALEX visits, FUV and NUV were exposed simultaneously, yielding one FUV image and one NUV image of the same field; the source extraction was performed by the GALEX pipeline in the NUV and FUV images separately, then measurements in the two bands were merged to create the source catalog.  SEP ranges between 0. and 6.986\as in the sample\footnote{However,  in the individual visits where one of the two detectors was not exposed (mostly FUV), SEP=-999.0 and only the magnitude from the recorded image is recorded in the GALEX Casjobs database, the other magnitude has a value of -999.0}.  We eliminate 1,259 sources with SEP $>$3.0\as , reducing the analysis sample to 69,745 sources\footnote{eliminating sources with SEP $>$3\as is  a very conservative limit; a cut at ``SEP''$>$5\as will only eliminate 260 sources. For this type of analysis, we choose a robust, albeit reduced,  sample rather than a numerically larger sample possibly contaminated by lower quality data}, with most values $\leq$2\as (Figure \ref{f_SEP_FWHM}, left panel).     Based on the histogram (right panel) in  Figure \ref{f_SEP_FWHM}, we further restrict the analysis sample to sources with NUV\_FWHM\_IMAGE and FUV\_FWHM\_IMAGE  $\leq$8~pxl (i.e., 12\as ; a GALEX virtual pxl is 1.5\as): 48,560 sources. If we retained sources with NUV\_FWHM\_IMAGE and FUV\_FWHM\_IMAGE  $\leq$10.pxl (15\as), the analysis sample would contain 59,374 sources. Most of the sources eliminated with these criteria can be recovered with custom photometry in the individual GALEX observations, but for the present bulk analysis we prefer a reliable UV photometry at the expense of reducing the sample numerically.   
 
 The reason why we find some GALEX sources measured by the pipeline as more  extended than the instrumental PSF among sources that the SDSS has treated as point-like  (taking the pipeline-choosen ``best''-mag measurements FUV\_MAG and NUV\_MAG), could arise from the very different morphology that some sources may have in UV $versus$ optical wavelengths (e.g., \cite{bia14}) and from the lower GALEX spatial resolution, whereby two or more very close sources may not be resolved by the GALEX pipeline but are resolved at the SDSS almost 3$\times$ better resolution. In fact, if we examine separately NUV\_FWHM\_IMAGE  and FUV\_FWHM\_IMAGE, there are more NUV sources with  NUV\_FWHM\_IMAGE larger than 10pxl than there are FUV ones (8,300 $vs$ 3,529). The difference can be explained by the NUV sources being more numerous (typically $\sim$10 times), and therefore more crowded than FUV-detected sources, that are more rare and sparse, causing the pipeline to merge close NUV sources in crowded fields while resolving the rarer sources in the FUV image. The point-spread-function (psf) is also slightly broader in NUV than in FUV, especially for very bright sources. Among the sources with   NUV\_FWHM\_IMAGE$\geq$10\as, 16.3\% have multiple SDSS matches to the GALEX {\it GUVcat} sources, while in the whole point-like  sample only 7.7\% have multiple SDSS matches. 
 Most of these sources can be recovered by custom photometry (e.g., \citet{demartino2420,bia14}) or in some cases by using small aperture photometry (also included in the catalog) and applying the aperture correction from \citet{morrissey07,alexbia}, instead of using the ``best'' pipeline measurements that treated the source as extended. We keep in the master catalog all sources, and all the existing magnitude measurements, so that interested users know that GALEX imaging is available, and  quality photometry can be recovered, for sources of particular  interest; the appropriate method may vary, according to object type and purpose. In this work we perform a statistical analysis of a large sample; therefore, we must use a consistent set of measurements, and restrict the analysis to the subsample where these standardized  measurements are expected to be reliable. 

All the 25 sources that have no visit-level detection within 3\as of the merged-catalog position in the full point-like sample, are eliminated by the above cuts in this culled sample (48,560 sources).  
 
One source,  GALEX\_ID=6378832251021428138,  has SDSS measurements only in the r-band, although it is rather bright at UV wavelengths: FUV\_MAG=18.14$\pm$0.076 and NUV\_MAG=18.62$\pm$0.045~ABmag.  The  FUV-NUV=$-$0.476$\pm$0.089 color makes it a candidate for a hot isolated star.  The Simbad database yields only one match with a known source, SDSS J134459.14-011038.2, classified as a Blue straggler. We do not eliminate this source for now, although its SDSS SED is incomplete\footnote{Another source, GALEX\_ID=6375841651334581457, does not have a SDSS g-band measurement, but it does have measured magnitudes in all other filters. The only known counterpart listed in Simbad for this source is Gaia~DR2~4209403408511502976, classified with Simbad-type ``Hsd\_Candidate''.  This source is not included in the culled sample}.

 Finally, for a number of sources the SDSS magnitude is saturated in one or more filters, which significantly alters the SED shape, hampering the SED model-analysis. After discarding sources with saturation flag in {\it g, r, i}, the analysis sample is reduced to 46,818 sources. Excluding also sources saturated in the z-band, there are 46,744 sources left. As many as 1234 sources are saturated in SDSS $g$-band, and these might be very interesting sources for our objectives. However, we eliminate  from the analysis in this work all sources with saturation in any of the SDSS bands $g, r, i,  z$; they might be recovered in future works with additional complementary optical data. The analysis sample is thus reduced to 46,744 sources with no SDSS saturation flags.

  Of the 46,744 
sources in the culled analysis sample, 5,891 ($\sim$12.6\%) 
have no Gaia match, 40,853 (87.4\%) have a match in Gaia DR3, as defined in Section \ref{s_gaia}, of which 38,728 have a unique Gaia counterpart (MMRANK\_GAIA =0) and only 2,125 have additional Gaia sources within the match radius (MMRANK\_GAIA =1)\footnote{we recall that only the primary match is retained in the master catalog, to preserve a unique source list, while the additional matches can be found in the full match result files, listed in Section \ref{s_cats}}. 
Among the 40,853 primary Gaia matches, 37,411 (80\%) have a parallax value in DR3; but as many as 3,885 (9.5\%) have a negative parallax value; of the 33,526 (82\% of the Gaia matches) Gaia counterparts with a parallax measurement $>$0mas (and PARALLAX\_ERROR is also positive),  15,586 (38\% of the Gaia matches)  
have  a parallax error better than 20\%.  This subsample is particularly valuable because, thanks to the availability of a good direct distance measurement, absolute stellar parameters such as Radius and \Lbol can also be derived from the stellar SEDs (Section \ref{s_sedfit}). Not surprisingly, the fraction of  Gaia matches with good  measurements is higher in the culled sample than in the overall sample (Sections \ref{s_sample} and \ref{s_gaia}).

About 58\% (27,440 sources) of the analysis sample have a Simbad match, all of which have  SIMBAD\_MAIN\_TYPE\_NEAREST not null. Among
the Simbad matches, SIMBAD\_MAIN\_TYPE\_NEAREST is ``Star'' for 1,116 sources and ``WD*\_Candidate'' for 12,178 sources (26\% of the sample). 
 Only a few cases have SIMBAD\_MAIN\_TYPE\_NEAREST indicating an  extra-Galactic object: 65 have type =  'Galaxy', and from zero to a few sources have other extra-Galactic types, with the exception of the SIMBAD\_MAIN\_TYPE\_NEAREST = QSO (1053) and = Seyfert\_1 (581); such contamination will be discussed in the following sections (see also \citet{bia09qso,bia11a}).

We have strived in the selection so far to increase the purity of the sample without giving weight to the quality of the SDSS photometry, lest  interesting UV-identified hot WDs might be eliminated because they have poor SDSS measurements.  We examine now the error distribution of the SDSS magnitudes, that is relevant for the analysis that follows and in the interpretation of the results. 

While the GALEX FUV and NUV magnitudes have a maximum error of 0.3~mag by construction of the database, we have not applied  so far any error cut to the SDSS counterparts. SDSS photometric errors are $>$0.3mag  in g~/~r~/~i~/~z for  354 / 810 / 1833 / 10542 SDSS counterparts (the distribution reflecting probably our selection of hot stellar sources), with a few sources in each band having error up to a few hundred mags.  But most sources that  have a large error in one band do not necessarily have large errors in other filters. As far as there are a few good colors, the SED can still be usefully analyzed, because each photometric band is given a weight inversely proportional to its error, for example in the SED fitting procedure with model colors. Therefore, we do not eliminate these sources from the online master catalog, as it can be used by others in other ways, and augmented by other data.  But we note here  that 233 SDSS counterparts have error  $>$0.3mag in all $g, r, i, z$ bands, 268 SDSS counterparts have error  $>$0.3mag  in  $g, r, i$, and 296 have error  $>$0.3mag in SDSS  $g$ and $r$.

For the following analysis, which involves both GALEX and SDSS measurements, we further restrict the sample by imposing error cuts of 0.2mag in all SDSS bands except for SDSS  $z$, resulting in a total of 35,294 sources. We used the SDSS pipeline ``PSFMAG'' measurements.  
Of the 35,294 sources, 10,598 do not have a Simbad match within 5\as (both SIMBAD\_MATCH\_NEAREST and SIMBAD\_MAIN\_TYPE\_NEAREST are
'=='), while for the 13,846 known objects that  do not have a spectral
type classification the field SIMBAD\_SP\_TYPE\_NEAREST is set to '==' if there is no counterpart, and to  '-888.' if there is a known counterpart but no spectral type.

\begin{figure}[!h]
\centering
\includegraphics[scale=0.11]{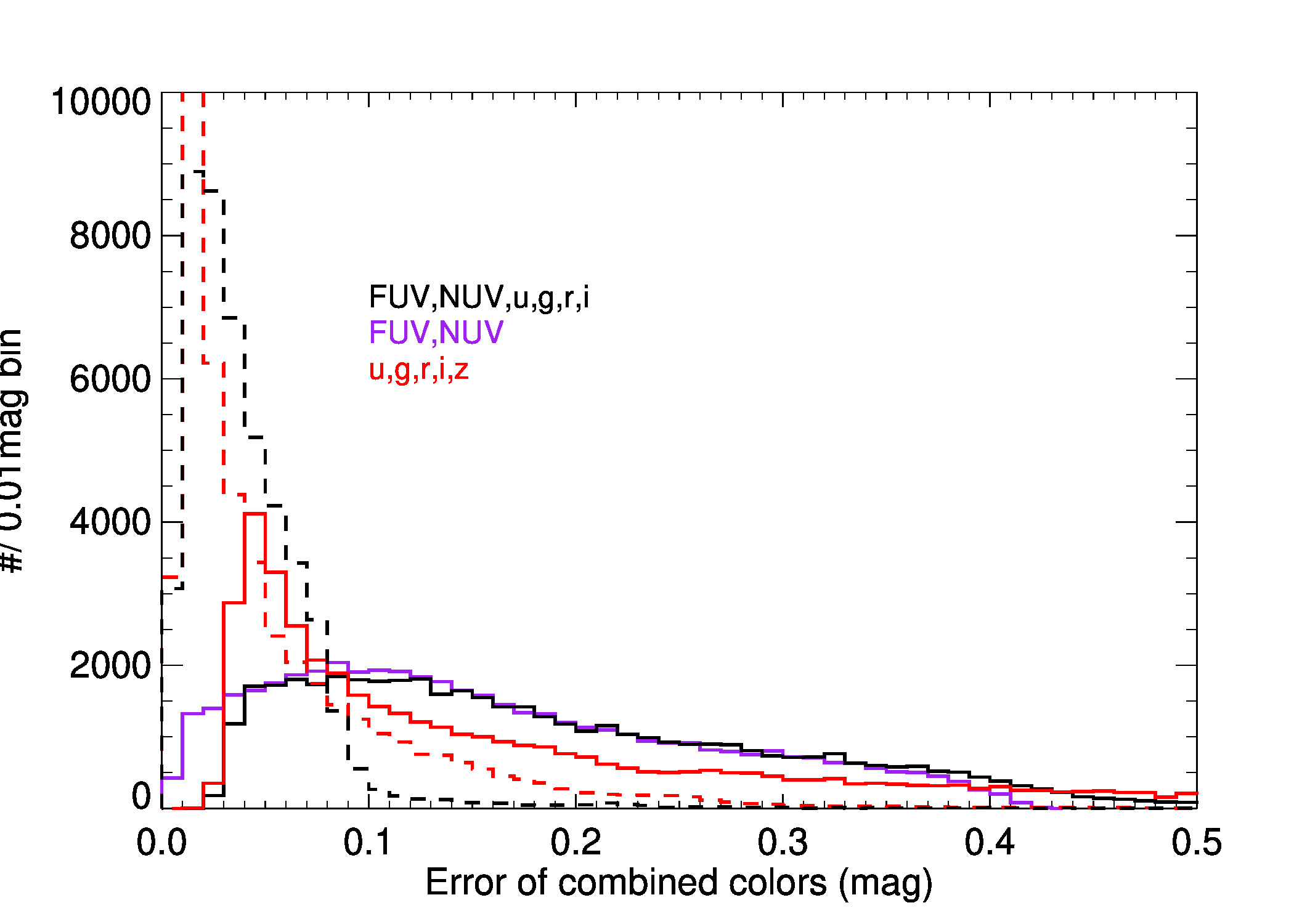}
\includegraphics[scale=0.11]{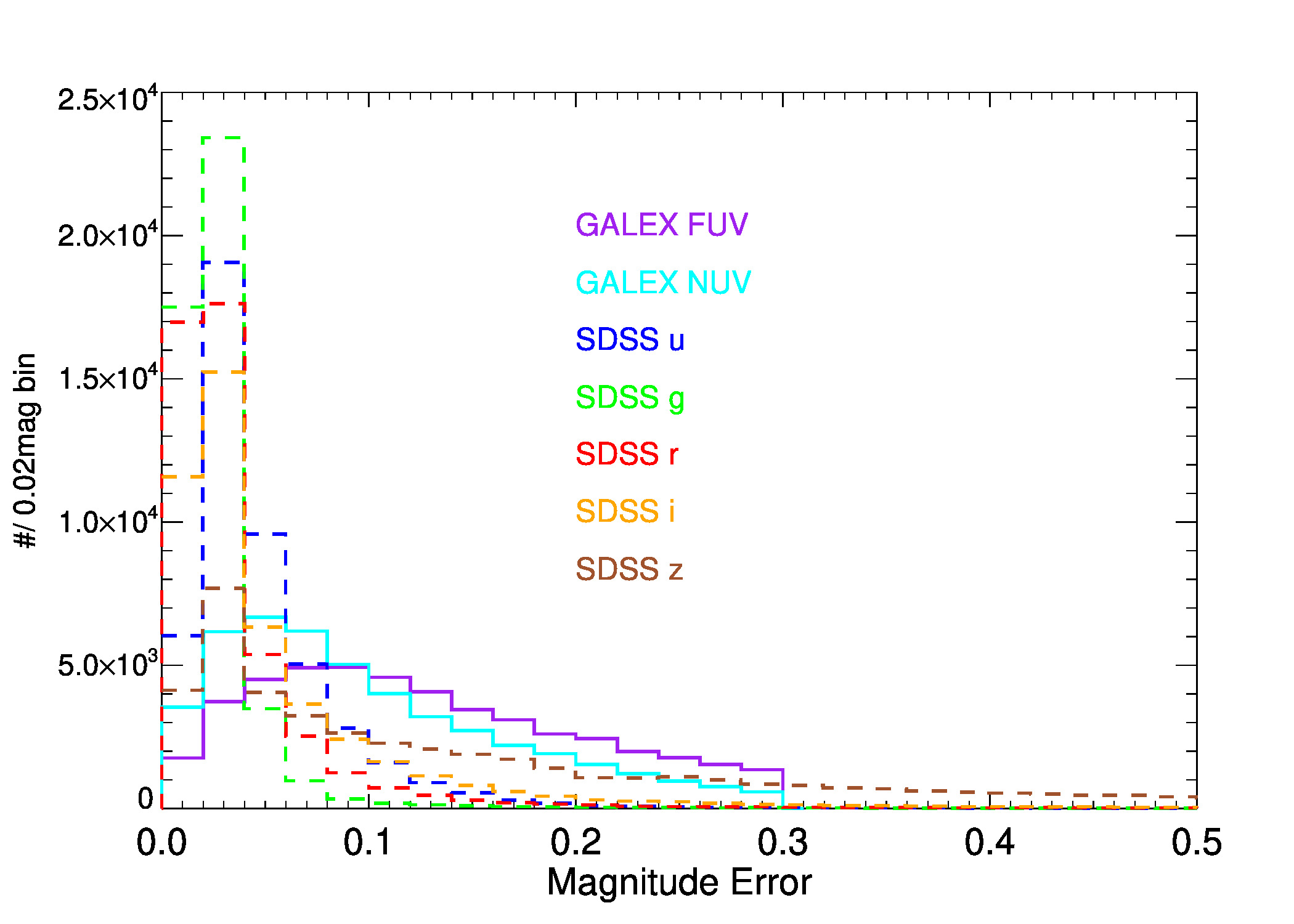}
\caption{
\small
{\it Left:} Distribution of combined errors in the culled analysis sample for the FUV-NUV color (purple), for FUV,NUV and SDSS {\it u, g, r, i} (black), and for the five SDSS bands {\it u, g, r, i, z} combined (red). The dashed histograms are the errors combined and divided by the number of colors, i.e. 1 for FUV-NUV, 5 and 4 for the other two combinations. 
 {\it Right:} Histogram of photometric errors in individual filters.  For the majority of sources, the SDSS photometric errors are smaller than the GALEX photometric errors; therefore, the SDSS magnitudes have more relative weight for example in driving the SED fitting results. However, a number of sources have SDSS photometric errors up to $\sim$10$^2$mag. 
\label{f_errcolors} }
\end{figure}

\subsection{Identification of Single and Binary Candidates from color-color Diagrams}
\label{s_binaccd}

In view of the growing evidence 
 that the majority of massive stars are formed in binaries  and that $\sim$70\% of these binaries interact and 20-30\% possibly  merge before reaching their final evolutionary stage 
(e.g., \citet{Sanaetal2014,Sanaetal2017,MoeDiStefano2017,abadie2010,patrick19,patrick20})
and - on the other hand - of the unique sensitivity  of our UV catalog to identify hot WDs, the evolved descendant of intermediate-mass stars, it is interesting to search in the present source catalog, which consists essentially of hot WDs (mostly) or SDs,  for candidate binaries including a hot evolved star and a cooler, less evolved companion. The purpose is to increase the statistical samples of such binaries, that are elusive in databases at longer wavelengths. A large, unbiased sample can support a variety of follow-up studies to clarify binary evolution (such as, for example, the intial-final mass relation). Here we perfom bulk color analysis  to estimate the binary fraction of these evolved stellar objects, and 
 to infer - by comparison with the fraction of unevolved binaries, measured in several studies by others - the probability that binaries merge
 before the post-AGB stage. 

\begin{figure}[!h]
  \includegraphics[angle=0,width=15.5cm]{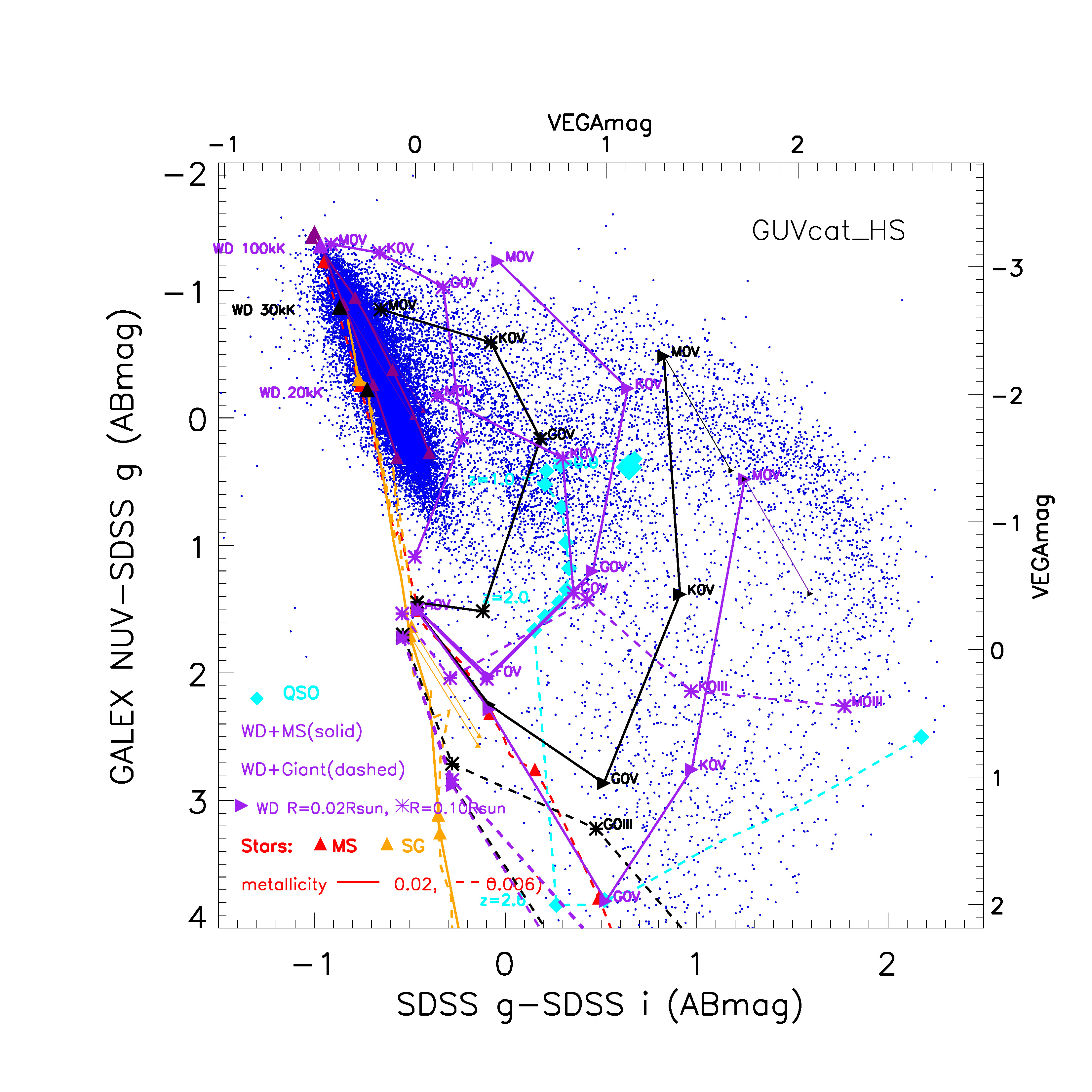} 
  \vskip -1.cm
   \caption{The sources (blue dots) are shown in color-color space,  along with model colors for hot compact stars (\logg = 7 and 9, \Teff = 100,000K to 15,000K, purple lines), main sequence stars and supergiants (red and yellow sequences, respectively), sample binaries consisting of a WD  of \Teff=100,000K, 30,000K, and 20,000K (purple, black, purple respectively), radius=0.02\Rsun and 0.1\Rsun, and a main sequence or giant star companion of representative spectral types. The QSO locus is also shown (cyan; the large diamond marks redshift=0). 
     Low-redshift QSOs 
     occupy a locus approximately in the center of this plot, which overlaps with some types of stellar binaries. 
     Model colors are unreddened; thin arrows shown on some models indicate \ebv=0.2mag reddening.
 \label{f_ccwide} }
\end{figure}

In color-color diagrams combining the GALEX and SDSS photometric bands, the single hot WDs populate a sequence well separated by other classes of astrophysical objects, as shown in Figure \ref{f_ccwide} (see also \citet{guvmatch} - their Figures 4 and 5, \citet{bia09,bia11a}). Also, binary hot WDs  with companions cooler than spectral type $\approx$A or $\approx$F  (main sequence or giants) can  be separated from single hot WDs and from other astrophysical objects, with the possible exception of some rare QSOs that have typical optical colors but anomalous FUV-NUV colors, due to enhanced \Lya~ emission in the low-redshift sample ($\sim$0.1 to 1.47) and to a dust phase for a smaller sample at redshift$\sim$2 \citep{bia09qso}. These QSOs, discovered by \citet{bia09qso}, represent a  small fraction of the QSO population ($\lesssim$5\%, according to the  estimate by \citet{bia09qso}). Most QSOs have typically fainter UV magnitudes than the present GALEX stellar sample, which is extracted from the AIS\footnote{AIS has a typical limiting depth of FUV=19.9 and NUV=20.8~ABmag \citep{bia17guvcat,morrissey07,bia09} }.

As mentioned in the previous section, for this analysis, which involves both GALEX and SDSS measurements, we  restrict the sample as culled in Section \ref{s_culling} by imposing error cuts of 0.2mag in all SDSS bands except for SDSS  $z$, resulting in a total of 35,294 sources. We used the SDSS pipeline ``PSFMAG'' measurements.

\begin{figure}[!h]
  \includegraphics[angle=0,width=17.cm]{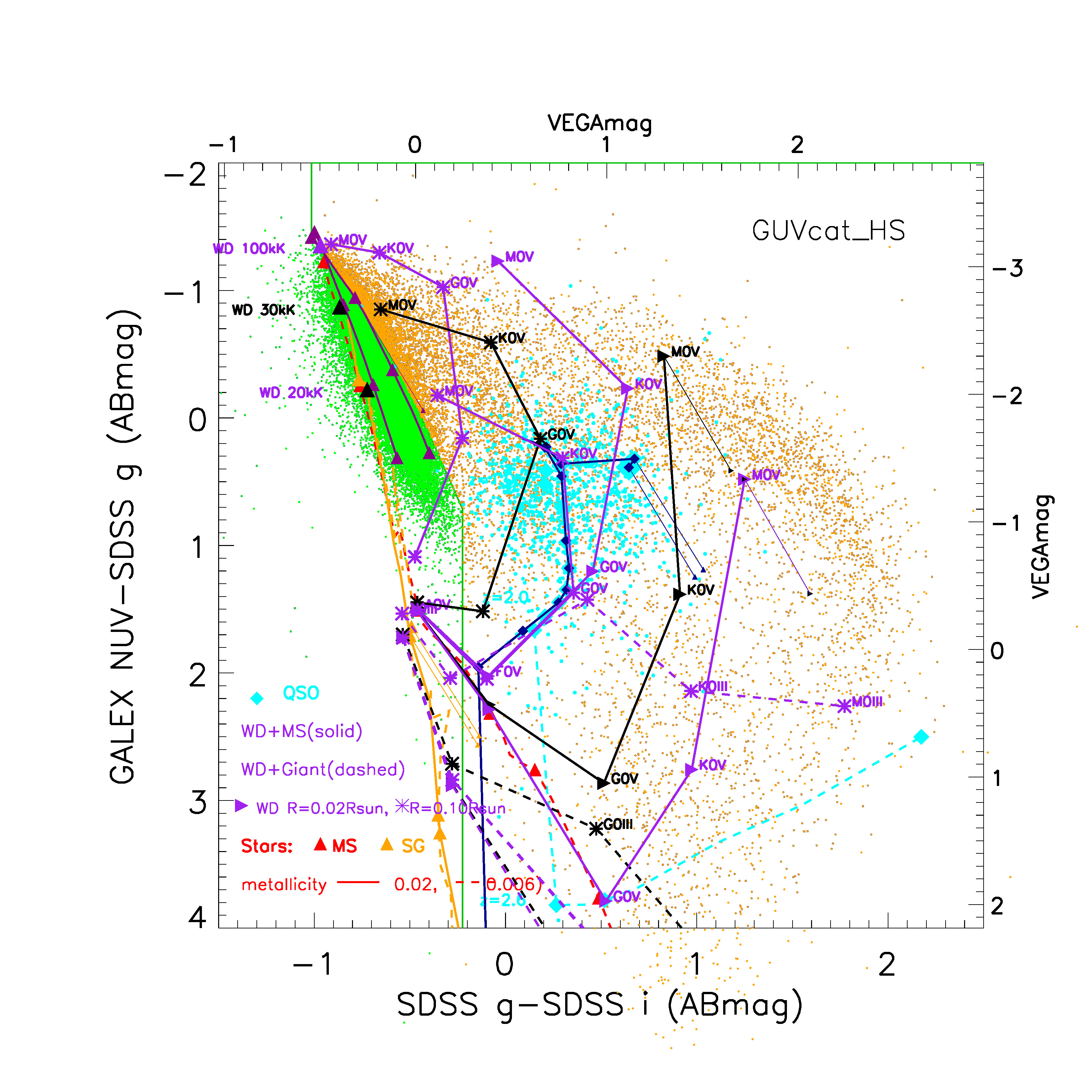}
   \vskip -1.3cm
  \caption{Same as in Figure \ref{f_ccwide}; a green line shows the    adopted separation between candidate single (green dots) and
    binary (orange dots) hot evolved stars. Sources with Simbad type =  QSO or AGN or Seyfert are overplotted with cyan dots, of larger
    size than the other sources for visibility, but they are only 941  sources out of the 35,294 total analysis sample; they occupy
    exactly the locus of the QSO average-template colors for redshift$\sim$0.5$-$1.5 (cyan for the standard QSO template, dark
    blue for the \Lya-enhanced template of \citet{bia09qso}). Triangles on the purple single WD sequences (\logg=7 and 9) mark \Teff =200kK, 30kK, 20kK, 15kK.
 \label{f_ccsel} }
\end{figure}

\begin{figure}[!h]
  \includegraphics[angle=0,width=14.5cm]{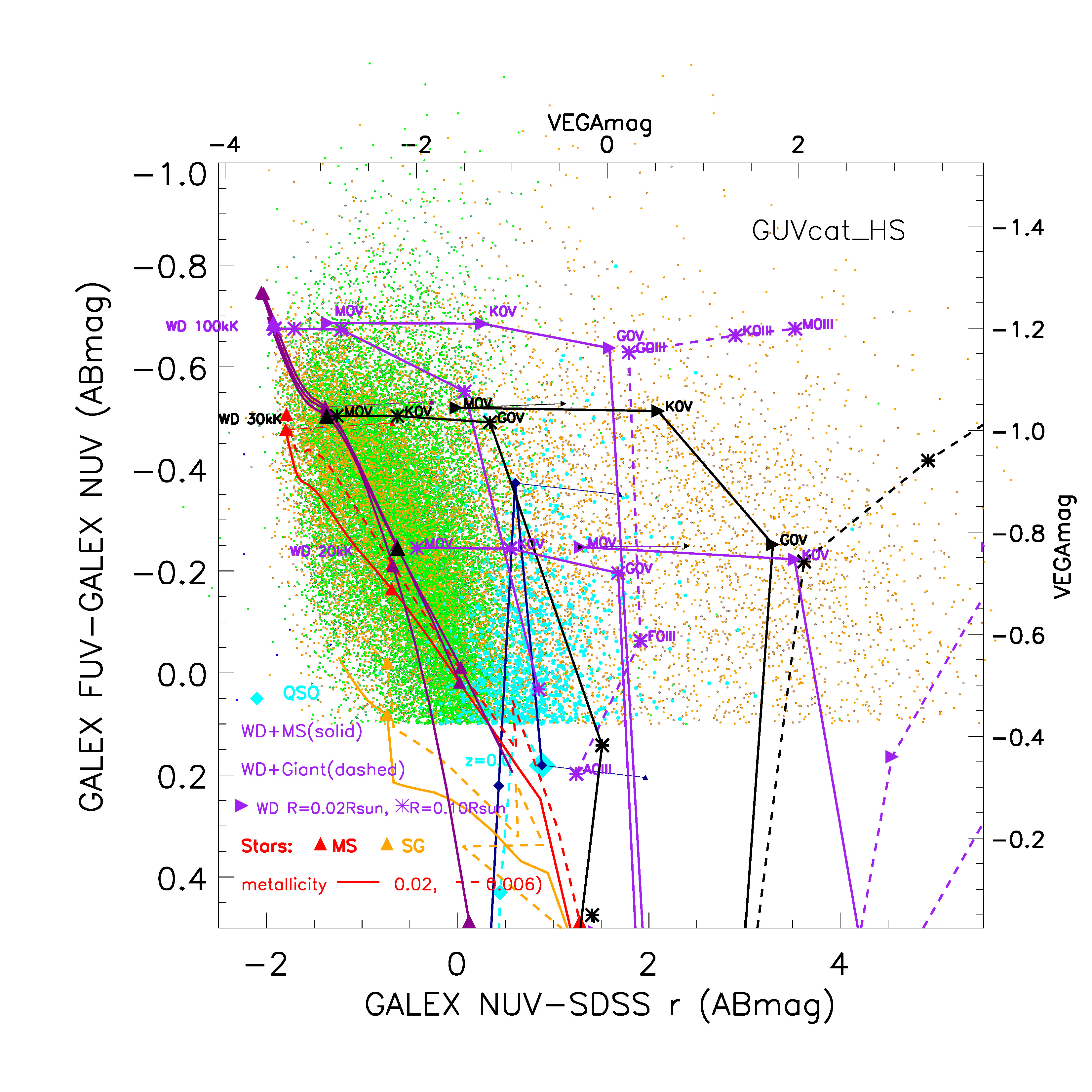}
  \vskip -1.2cm
\caption{\small{FUV-NUV $versus$ NUV-r. Data and model colors as in the previous figures. The sample is restricted to FUV-NUV$\leq$0.1mag, but a wider Y-axis range is plotted to  show also types of single stars and binaries excluded by the color cut.  NUV-r separates  main-sequence and giant stars from the WDs (\logg= 6, 7 and 9 shown) more than in previous plots. In FUV-NUV,  standard-template QSO colors (cyan) differ significantly from \Lya-enhanced QSOs (dark blue, \citet{bia09qso}). 
 Known QSOs+AGN+Seyfert, marked with large cyan dots,  mostly cluster around FUV-NUV$\gtrsim$0, where the template colors predicts them to be.  Reddening arrows for \ebv=0.2, shown on some models,
 are nearly horizontal, because  FUV-NUV is essentially reddening free. The spread of the observed QSOs 
 around the un-reddened template colors is due to photometric uncertainties and variations around an average template, the larger spread in the extinction direction is due to reddening. At FUV-NUV$\leq$-0.3mag, 
 the contamination  from QSOs and B supergiants is negligible.  Single-star and binary candidates selected in Figure   
 \ref{f_ccsel} are marked with green and orange dots, respectively. 
 }  \label{f_ccselFN}  }
\end{figure}

Figure \ref{f_ccwide} shows a combination of one GALEX band, NUV, with SDSS $g$ and $i$; the data-points (blue dots) are shown as well as model-colors constructed from model atmospheres  for main sequence and supergiants, and for compact stars with \logg = 7 and 9 (models described in \citet{guvmatch,bia24grids}). As previously noted by \citet{guvmatch} and \citet{bia11a},  Figure \ref{f_ccwide} shows that the hot-star sequences with \logg = 7 and 9 enclose the majority of data-points for the current sample. Because of the FUV-NUV$\leq$0.1mag cut, all single stars cooler than spectral type $\sim$A are eliminated in this sample, while they can be seen to populate well the main-sequence and supergiant \Teff color-sequences in the full sample of \citet{guvmatch}. Only the locus  of the hottest massive early-type stars is adjacent to the hot WD locus at the hottest \teff, 
and these would be hard to distinguish from the hot WDs given the photometric errors on the colors, except for  the sources with a Gaia distance, for which  a radius can be derived from  SED analysis, see Section \ref{s_sedfit}.  However,  massive stars (O and B types) are  statistically very rare, because the IMF is skewed towards low masses and because massive stars evolve very fast, and for the purpose of source counts their numbers hardly affect the hot-WD population.

   Figure \ref{f_ccwide} shows also composite model colors for binaries composed of a hot WD (examples for three WD \Teff values  are shown: \Teff=100,000K, 30,000K and 20,000K, each for two extreme values of radius, R$_{WD}$=0.02 and 0.1\Rsun) and cooler, less evolved companions of representative spectral types, both main sequence and giants. Although only a few examples are shown to avoid crowding the plot,  these binary model-color sequences span all the color-color space occupied by the data-points outside the single-WD color locus.  The model colors are shown for unreddened stars, but  extinction arrows are drawn on a few sample model colors.  The two-colors combination in Figure \ref{f_ccwide} also shows that binaries with companions hotter than $\sim$A (depending on the ratio of radii within the stellar pair) are either excluded by our sample's UV color restriction, or would hardly be distinguishable from a single A-type (and earlier) star.  Comparing this figure with Figure 4 of \citet{guvmatch} we also note that a large number of sources in their sample occupy the color-color locus covered by galaxy models: those data-points are all extended sources (black points in \citet{guvmatch} figures), which are excluded from this sample of point-like sources.  The only extra-Galactic objects that may intrude the point-like sources are QSOs or AGNs, with a central light peak dominating the faint underlying galaxy, at low red-shift because of the FUV-NUV$\leq$0.1mag cut. 

In Figures \ref{f_ccwide} to  \ref{f_ccselFN} sequences of QSO models colors are also shown from the ``average QSO'' template (cyan symbols), and in Figures  \ref{f_ccsel} and  \ref{f_ccselFN} also from the enhanced-\Lya~ template (dark blue) of  \citet{bia09qso}. QSO model symbols (diamonds) mark redshift = 0, 0.2,    0.4,     0.6,      1.0,      1.2,      1.4,      1.6,      1.8,      2.0,      2.2,
      2.4,      2.6,      3.0 and      4.0; some values are labelled. The redshift=0 model locus is shown with a large diamond. As expected from the GALEX pass-bands, and from our UV-color selection, only low-redshift QSOs are included. 
  In  Figure \ref{f_ccsel} we mark with large cyan dots the 954 sources classified as extra-Galactic (QSO or AGN or Seyfert)  in Simbad.

Based on the loci defined by the model colors,
 we define a separation between single and binary stellar candidates, shown with a green line in Figure    \ref{f_ccsel}. The boundary follows the model sequence of the WDs with \logg=9 (the single-WD sequence closest to the binary locus) with minimal extinction (\ebv=0.1mag) and   
extends in NUV-$g$ to redder colors, because single stars cooler than \Teff$\sim$15,000K are excluded from the sample by the FUV-NUV color limit, as is also evident from the color-color figures, while binaries - owing to the presence of the hot WD-  can occupy redder colors in the diagram. 
  In Figure \ref{f_ccsel} the binary candidates are marked with orange dots, and the single hot-star candidates with green dots.   We track these candidates with the same color-coding in Figure \ref{f_ccselFN}, showing the FUV-NUV color in the Y-axis. Note in this plot that the extinction arrows are nearly horizontal, because the broad FUV-NUV color is essentially reddening-free for Milky-Way type extinction (Section \ref{s_sample}).  The single-hot-star candidates (green dots) still occupy a defined locus, although with a wider spread in NUV-{\it g} (X-axis) compared to the previous color combination, due also to the larger reddening effect because the color spans a wider wavelength range (note also the different X-axis range in the figures). For this reason, the initial selection was made in the color combination of Figure \ref{f_ccsel}. Instead, the binary candidates, the orange points, occupy the whole width of the diagram, including the single-star locus, consistently with the binary model predictions.   In FUV-NUV, the \Lya ~enhanced QSOs at low redshift (dark blue model colors) stand out well separated from the standard QSO template (cyan).  The overplotted known QSOs + AGNs + Seyfert  (large cyan dots)  occupy a well defined locus in both Figures \ref{f_ccsel} and \ref{f_ccselFN}, overlapping with the locus of certain binary combinations.  Figure \ref{f_ccselFN} also shows that the hottest sources, with FUV-NUV $\leq$-0.3mag, are rather free of QSO contamination.

From the two-color separation illustrated in Figure \ref{f_ccsel}, we count  
 22,848$\pm$$^{1267}_{3853}$ sources in the ``single-hot-star'' locus (green dots) and 
 12,404$\pm$$^{1871}_{1267}$  binary candidates (orange dots).   The uncertainty in the counts is estimated by applying to each color  the combined photometric errors of the two band measurements, and recounting the numbers in the single and binary loci  adding and subtracting  the  error to the nominal colors of each source.   
As discussed above, the stellar-binaries locus contains a small number of intruding QSOs;  therefore,
 the binary candidates count of 12,404 is an upper limit.  We estimate a correction based on the known objects, and on the Gaia-match statistics. 
Excluding from the binary candidates the 954 sources with an extra-Galactic classification in Simbad, 
 the binary fraction would be  \fbina= 0.50\%. The known QSOs or AGN or Seyfert are 7.7\% of the sources in the binary-defined locus.  But only a fraction of the sample has a Simbad match to a known object: 
 only 48.6\% (6,030) of the binary candidates have a Simbad identification (6,274, about 51\%, have SIMBAD\_MAIN\_TYPE~=~'==', i.e. are unknown objects). If the ratio of extra-Galactic over Galactic sources were the same in the un-classified half of the sample, the contamination by extra-Galactic objects would be twice the 7.7\% fraction of the known ones. With this correction, the binary fraction would be \fbina=0.46\%. On the other hand, because the most elusive astrophysical objects are the hot WDs, in surveys at optical wavelengths, assuming a contamination by extra-Galactic objects in the culled sample of binary candidates of 15\% is probably an overestimate.

\begin{figure}[!h]
\centering
\includegraphics[scale=0.35]{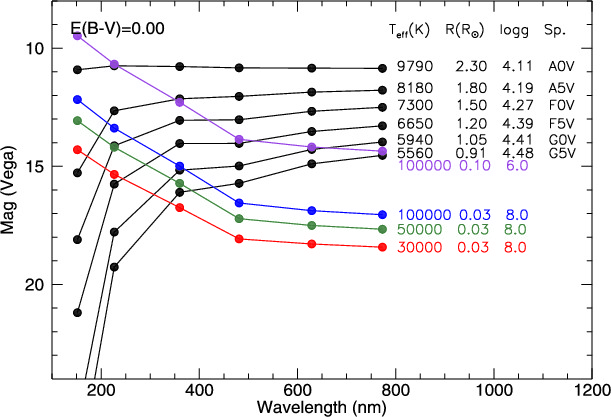}
\includegraphics[scale=0.35]{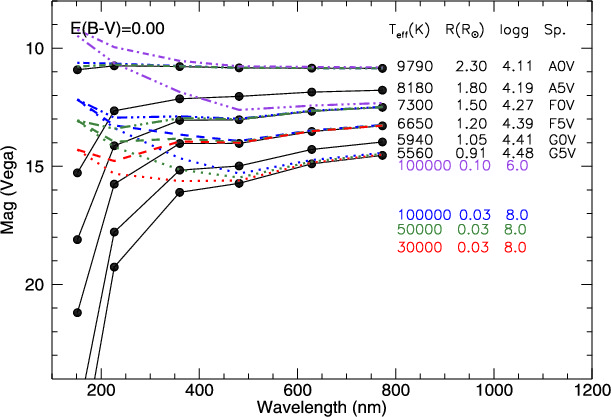}
\caption{
\small
{\it Left:} model SEDs in GALEX+SDSS filters, 
covering a range of stellar pairs targeted
by our study. Model mags for cool stars (Main Sequence (MS) case shown) in black, WD models 
color-coded (the stellar parameters of the 
purple model are appropriate for a central star of planetary nebula). {\it Right:} black dots 
are the single-star  MS models as in the left panel, 
dashed/dotted lines are  {\it composite} WD+MS model magnitudes for sample pairs. At the $\sim$4.5\as and 1.4\as resolution of GALEX and SDSS imaging, all pairs are unresolved and the SED looks like the dashed or dotted model-mag lines. 
The figure illustrates the power of FUV$-$optical catalogs to identify candidate binaries in a regime different from what
is accessible from optical data alone.
Model magnitudes are scaled to a distance of 1kpc, and to the stellar radii given in the legend. 
\label{f_SEDMSandWD_SDSS} }
\end{figure}

 We also recall that, with the presently available colors (FUV to near-IR), companions of types $\sim$A  and hotter  (and in some cases M0V and cooler, depending on the ratio of the stellar radii)  may escape the identification as binaries,  as shown also in Figure \ref{f_SEDMSandWD_SDSS}, where we plot the composite SED of GALEX+SDSS bands of a number of cool star types with hot WDs of representative \Teff and radii. 
 Single stars of these spectral types and later are excluded from the sample because of the FUV-NUV color cut, as shown in Figures \ref{f_ccwide} to \ref{f_ccselFN}. Therefore, they would be missed from both the binary- and the single-star counts. 
 The count of sources in the ``single hot star'' locus must be regarded as  an upper limit to the number of single WDs or SDs, because the SED of a stellar pair with very similar \Teff  will not be distinguishable from a single-star SED with the same \teff.  
Although  such ``identical twins'' may be rare cases among WDs, and the number of massive OB stars is also negligible, the counts of single hot WDs is in principle an upper limit.  With all these caveats, the fraction of binaries detectable as composite SED (i.e. not merged) is then  
\fbina$\gtrsim$50\%$-$46\%,
  less than the $\sim$75\% reported by most works for massive stars \citep{patrick19,Sanaetal2014,Sanaetal2017,MoeDiStefano2017}, but similar to the fraction reported for the lower mass range of the WD progenitors, from $>$80\% to 50\% in the mass range from 8 to 1\Msun~  according to \citet{moe2019}.

  If we consider the hottest stars among the sample, for example the sources  with FUV-NUV$\leq$-0.3mag (\Teff $\gtrsim$20,000-30,000K), the binary fraction is 57\% (10,122 single and 5,894 binary candidates including only 74 known objects classified as QSO or AGN or Seyfert) or  56.7\% if we assume twice the known extra-Galactic objects contamination. 

In sum, considering that the single-candidate counts are an upper limit, because they may contain identical-twin binaries,  and that  hot WDs with companions earlier than $\sim$A types are missed in binaries counts, 
the binary fraction for evolved stars is estimated as $\gtrsim$46\% for the whole sample (subtracting an assumed 15\% contamination by QSOs and AGNs), and 56.7\% for the hottest sample.    

\begin{figure}[!h]
\includegraphics[angle=0,width=13.cm]{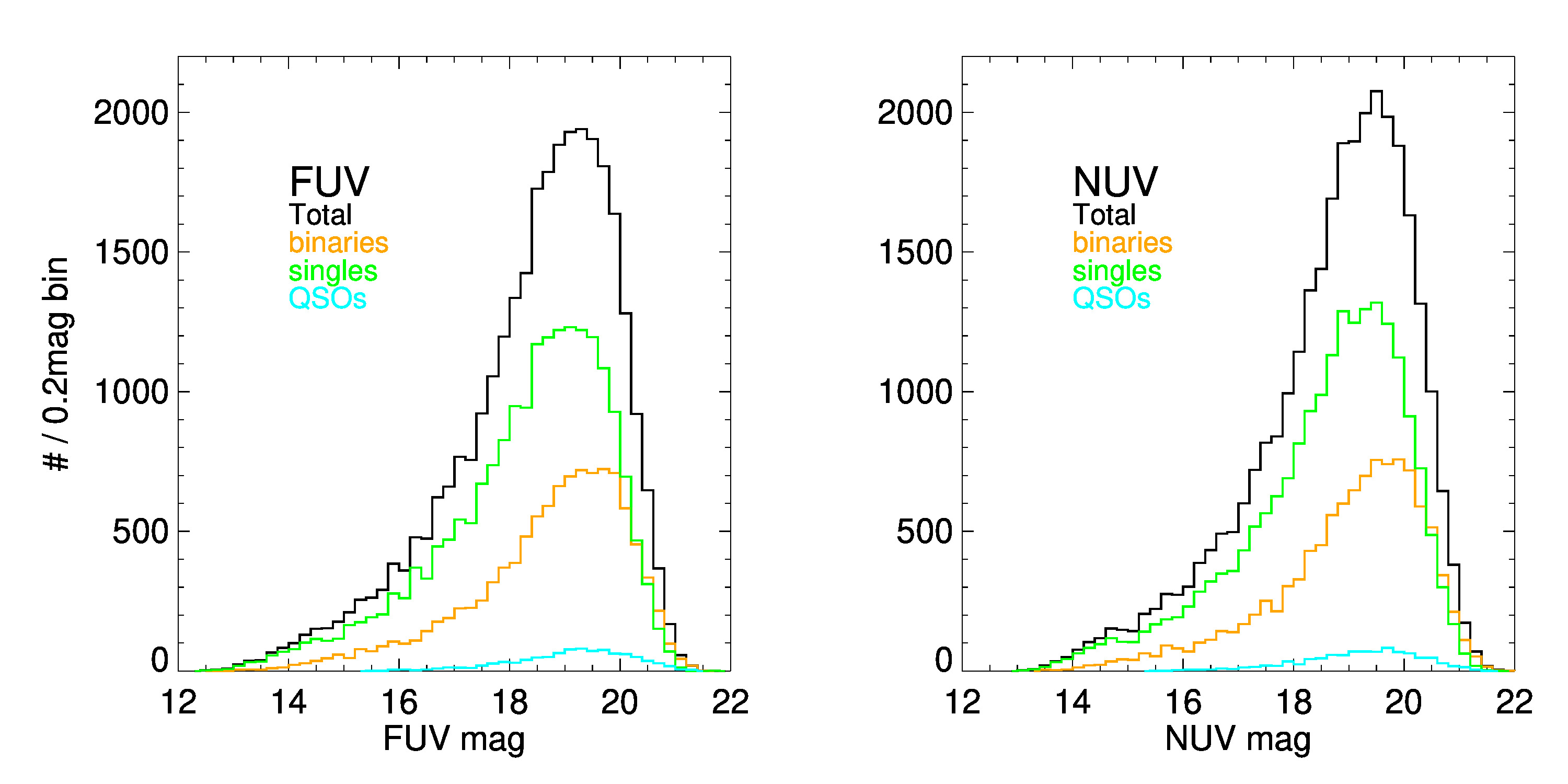}
\caption{Histogram of FUV and NUV magnitudes of the total clean sample (SDSS errors $\leq$0.2mag), and of the single (green) and binary (orange) hot-star candidates, and of the known sources classified as  QSOs or AGN or Seyfert (cyan). The contamination by the known extra-Galactic objects is minimal ($\leq$8\%, but only about half of the sample consists of known objects with classification). At the depth of the GALEX AIS, from which this sample is extracted, QSOs do not stand out for increasing at faint magnitudes, as they do in the deeper MIS survey \citep{bia11a}, where the number of extra-Galactic sources with respect to Milky Way stellar sources increases rapidly at fainter magnitudes. Therefore, in this sample, the QSO contamination in the binary locus cannot be eliminated  by their magnitude and colors. 
 \label{f_histoselmag} }
\end{figure}

\subsection{SED analysis to Characterize Binary and Single WD Candidates} 
\label{s_sedfit}

\begin{figure}[!h]
\vskip -1.cm
\includegraphics[angle=0,width=12.cm]{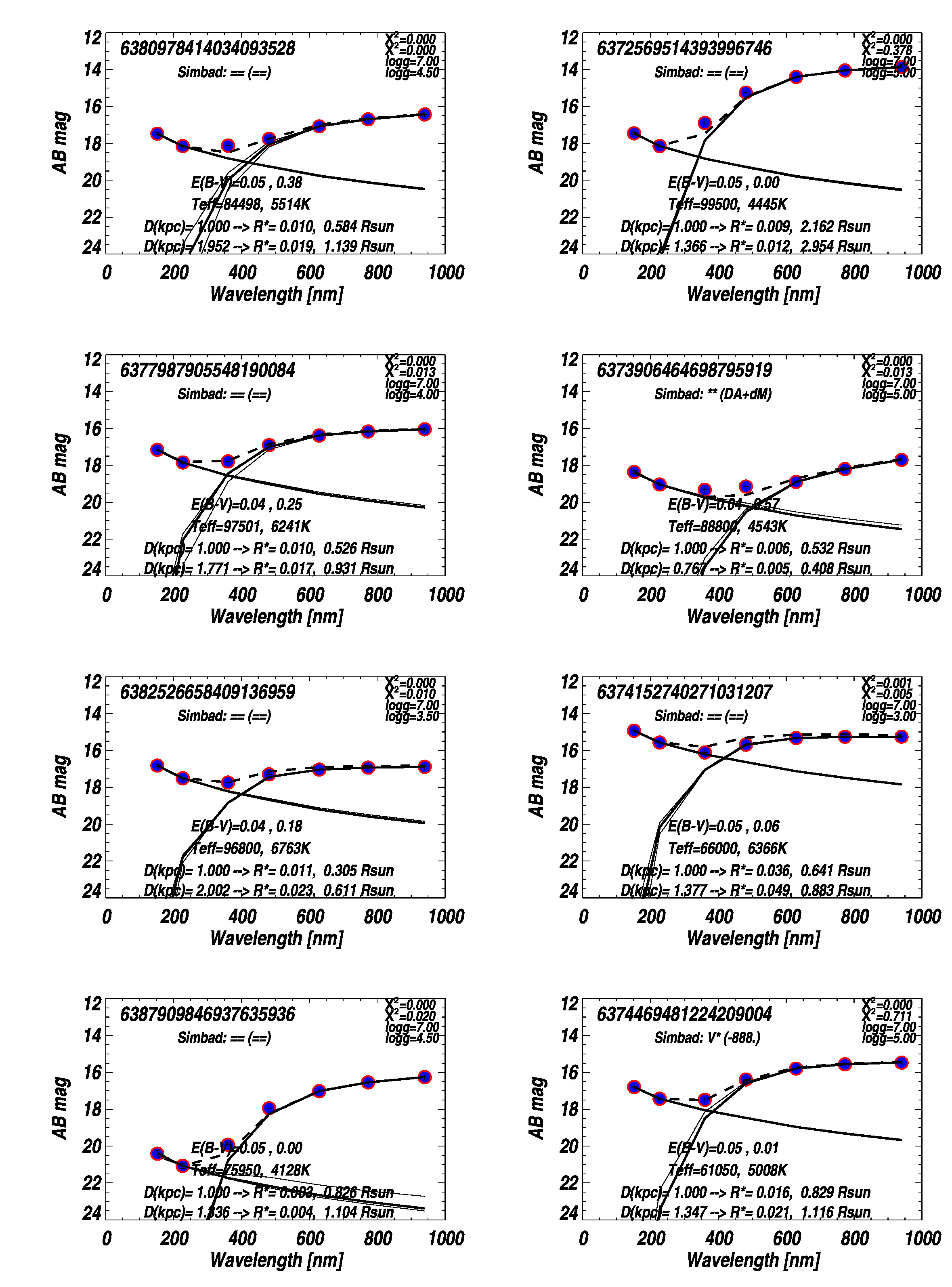}
\vskip -.5cm
\caption{Examples of two-component SED-fitting with stellar model magnitudes  (Kurucz for the cooler companion, Tlusty for the hot companion). The Simbad identification, and spectral type (in parentesis), are given when available. Resulting \Teff values are printed for both components, and of \Ebv (derived for the cool star, assumed =0.05mag for the hot star). Radii derived from scaling the best-fit model are given first for a reference distance of 1kpc, then rescaled for the distance derived from the Gaia parallax. Dots are observed magnitudes, the black thick line connects the best-fit model magnitudes, thin black lines show the acceptable range of solutions. 
  The dashed  line is the composite model SED from both components. 
 \label{f_SEDbina} }
\end{figure}

\begin{figure}[!h]
\vskip -1.cm
\includegraphics[angle=0,width=13.5cm]{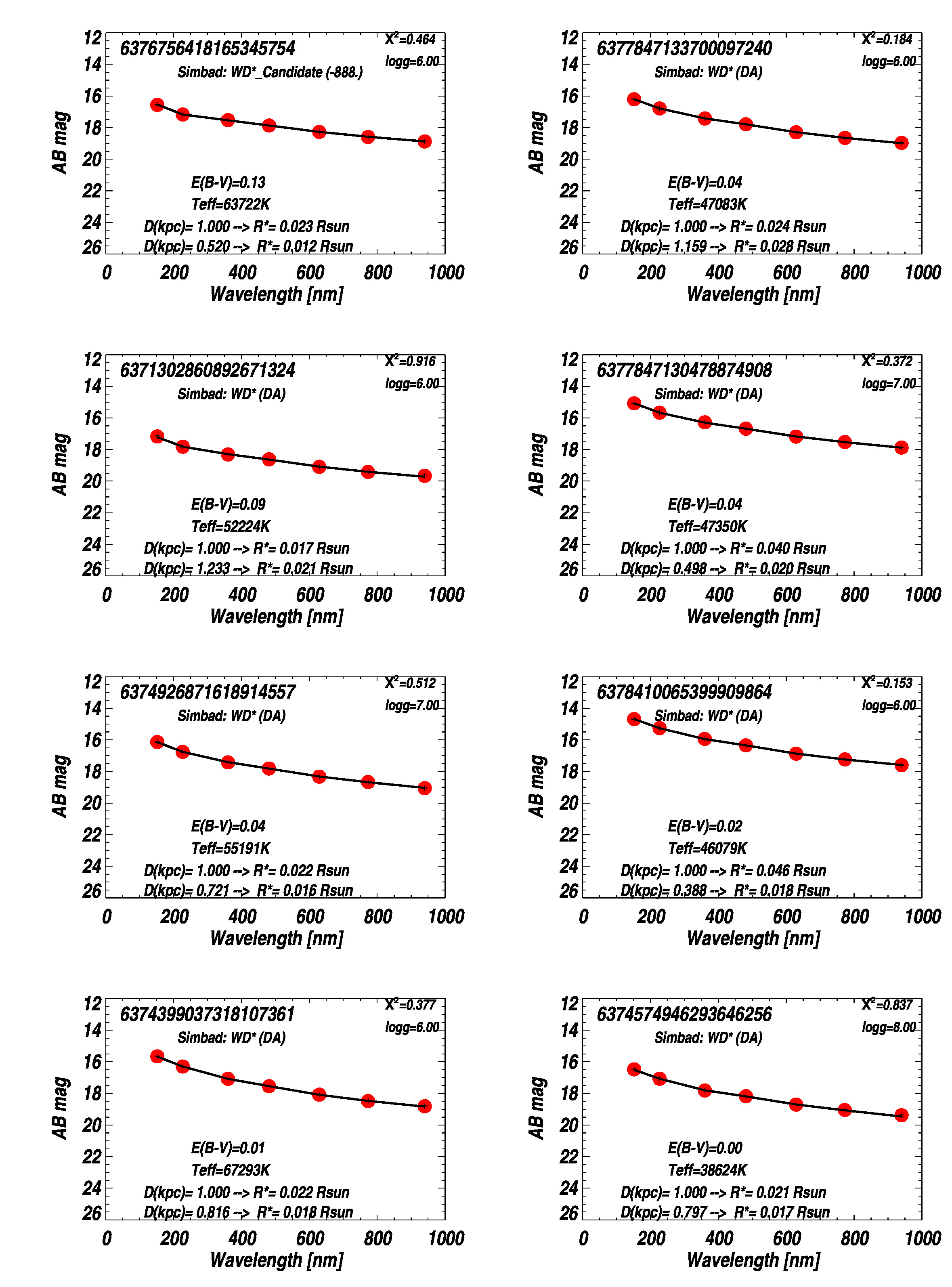}
\vskip -.5cm
\caption{Examples of  SED-fitting with Tlusty high gravity stellar models for single-star candidates with derived high \teff's. The Simbad identification, and spectral type in parentesis if available, are given.  The values of \Teff and  \Ebv (derived as free parameters, and the best-fit \logg) are printed;  the radius, derived from the best-fit-model scaling, is given for a reference distance of 1kpc as well as rescaled for the distance derived from the Gaia parallax. Symbols as in Figure \ref{f_SEDbina}.
 \label{f_SEDsingTlu} }
\end{figure}

\begin{figure}[!h]
\vskip -1.cm
\includegraphics[angle=0,width=12.cm]{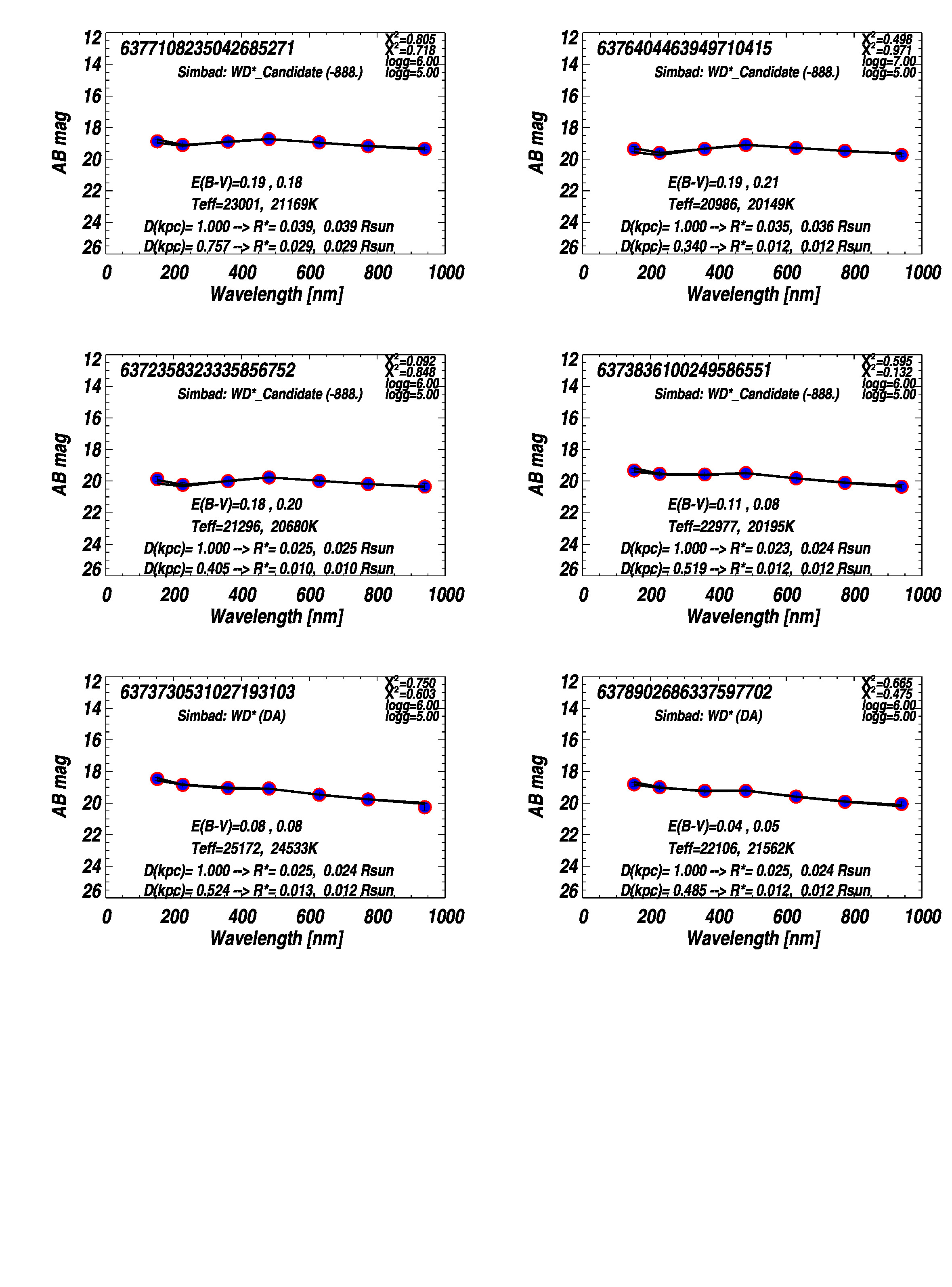} 
\vskip -3.5cm
\caption{Examples of  SED-fitting for single hot-star candidates with derived \Teff between 30kK and 20kK, from both Tlusty high gravity stellar models (\logg~ range 6$-$8) and Kurucz stellar models (\logg~ range 3$-$5). Resulting values, printed on the legend for both solutions, are generally consistent; the two best-fit model mag lines, both plotted with a black line, are indistinguishable on this scale. Most sources have a small radius, compatible with a compact evolved object; the solution from Kurucz model magnitudes finds the highest gravity available in the grid (\logg=5) as the best-fit case, but the actual gravity is likely higher in these examples, which is reflected in the discrepancy among the derived radii. 
\label{f_SEDsingTluKur}  }
\end{figure}

Using the color separation between single and binary hot stars from Section \ref{s_binaccd}, we performed preliminary SED-fitting of the UV$-$optical SED with the model grids of \citet{bia24grids}, from which the model colors shown in Figures \ref{f_ccwide} to \ref{f_ccselFN} are constructed. A full analysis will be the subject of another work; here we only show some examples  in Figures \ref{f_SEDbina}, \ref{f_SEDsingTlu} and \ref{f_SEDsingTluKur} to illustrate the potential use of this catalog.  When the 7-bands SED  can be fitted with a single-star model, both \Teff and \Ebv can be derived concurrently, because there are more independent colors available than free parameters; we performed the fitting with model grids of different \logg~ values (from 3.0 to 5 for Kurucz models, solar metallicity, and 6 to 8 for Tlusty pure-H models for the compact stellar objects). We fitted the SEDs with \Teff and \Ebv as free parameters (interpolating in the model grids), then choose the \logg~ value that yielded the lowest  best-fit  \Xtwo from all existing \logg~ values.  Once \teff, \Ebv and \logg~ are derived from the SED shape, scaling the best-fit model to the observed SED, accounting for the derived extinction, yields the stellar radius modulo distance, R$^2$/D$^2$. 
For the Gaia sources with a good parallax measurement (we choose a limit of parallax error $\leq$20\%), R can be scaled to an actual value, and \Lbol is then derived from R and \teff. 

 For the binary candidates, that have a composite SED, only the FUV-NUV color can be used  to derive \Teff of the hot component; therefore, only one parameter, \teff, can be derived independently.  We initially assumed \logg=7 (which is an average value for the sample, from Figure \ref{f_ccwide}), and a minimal reddening \ebv=0.05mag; this value turned out to be likely underestimated except at the high Galactic latitudes, because it yields many very small WD radii. Therefore, scaling the model of the appropriate \Teff to the model grid yields a lower limit to the WD radius, and - when a Gaia distance is available - to \lbol.  We use the redder SDSS magnitudes to derive \Teff and \Ebv for the cool companion, and constrain its \logg~ value, although these parameters are more uncertain than for the cases where all bands can be fitted as a single-star SED. With further iterations, the hot component of the stellar pair could be assumed to have the same \Ebv as derived from the cool companion, when the result is robust, and \Teff be derived from the FUV-NUV color with Tlusty models appropriately reddened.   Such more detailed analysis will be the subject of a follow-up work, where additional data will be combined. In this work the scope was limited to a bulk exploratory analysis, in the interest of releasing the catalog for possible exploitation by others. 
 
In Figures \ref{f_SEDbina} to  \ref{f_SEDsingTluKur} 
we show examples of preliminary SED model analysis, for binary and single hot-star candidates. Figure \ref{f_SEDbina} shows some of the binary WDs  with the hottest derived \teff:  most of these do not have a Simbad identification.  The reason is quite evident from the examples in Figure  \ref{f_SEDbina}: without the UV measurements, the optical-IR SED would be fitted with a single cooler star, and the presence of a hot secondary component would not be detectable, as it contributes negligible flux to the optical bands, unless the cool companion is also a subdwarf.  From the preliminary analysis we also noted that, for the binaries, the SED fitting results are very uncertain when the cooler companion's \Teff approaches 10,000K or hotter;  reliable fits are usually achived with companions cooler than 8,000K, as could also be guessed from the color-color plots in Figures \ref{f_ccwide} to \ref{f_ccselFN} and from the model SED examples in Figure \ref{f_SEDMSandWD_SDSS}.

~ \vskip -1.cm

~ \vskip -1.cm

~ \vskip -1.cm

\section{Discussion. The Binary Fraction of Hot Evolved Stars}
\label{s_binaf}

An interesting result from this work is the estimate of the fraction of binary  $vs$ single  candidates hot-compact stars in the   range of parameters where reliable counts can be derived from the present data. 
A wealth of studies has been recently devoted to estimate the binary fraction of massive stars  and their immediate descendants, the Wolf-Rayet (WR). For example, \citet{kobu14} infer a fraction of $\sim$55\% for binaries with orbital periods P$<$5000days, studying 48 massive systems in Cyg~OB2;  \citet{sanaetal2012} found \fbina(observed)=0.71 or \fbina(intrinsic)=0.69 from a sample of 71 O-type stars in six open clusters;  \fbina$\sim$0.4 was found from a compilation of 227 Galactic WR \citep{vanderhucht2001}), \fbina(observed)$\sim$0.36 from  
RV monitoring of 11 WN  \citep{dsilva23}, \fbina(observed)=0.44 or \fbina(intrinsic)=0.56 for 16 WNE \citep{dsilva22},  \fbina(observed)=0.58 
for 12 WC stars \citep{dsilva20} but \fbina(intrinsic)$\geq$0.72 according to \citet{dsilva22}, and \fbina=0.70 for LBVs \citep{mahy22binaLBV}.
The ensemble of such studies (and the list is not complete) showed that the binary fraction in massive stars is very high, with a significant number of main-sequence O-type stars also in hierarchical triple systems \citep{patrick19,Sanaetal2014,Sanaetal2017,MoeDiStefano2017,abadie2010}, and that many of these binaries exchange mass during their lifetime (71\% of all massive stars),  20\% to 30\% of which will merge  \citep{sanaetal2012}, based on the distribution of the orbital periods.  The cited studies are based on detailed observations of individual objects, including radial velocity (RV) monitoring of the binary systems to assess period distribution in order to correct for sample-selection biases; their only limitation is essentially the necessarily small numerical sample.
 The estimate of the number of very massive stars that undergo mass exchange, stripping, or merging with a close companion is relevant to explain for example the different classes of supernovae and their relative numbers (e.g., \citet{sanaetal2012} and references therein).  The progenitors of hot WDs are intermediate-mass stars (masses up to $\sim$8-9\Msun), i.e.  late-O $-$early~B type and later, although the frequent occurrence of mass exchanges within  binary pairs at early stages could stretch the zero-age mass range both ways, in case of significant mass transfer to a companion or mass accretion from a companion.
The estimated binary fraction for stars with initial masses between 8 and 1\Msun varies from $>$0.8 to 0.5 \citep{moe2019}.
The binary candidate overall fraction derived in this work, \fbina$\gtrsim$50\% or $\gtrsim$46\% (assuming a 7.7\% or 15\% QSO contamination), refers to our sample of hot compact objects (mostly WDs) with a detectable cooler (less evolved) companion $vs$ hot compact stars  with a single-star-like SED, i.e. it includes mostly pairs in which one object has evolved through the post-AGB phase.
The single stars of type $\gtrsim$A are excluded from the present sample by the FUV-NUV$\leq$0.1mag  limit.

Our identification of binary candidates is based on broad-band photometry, and, as discussed in previous sections, and as Figures \ref{f_ccwide} and \ref{f_SEDMSandWD_SDSS} illustrate, the probability of distinguishing a binary from a single star largely varies depending on the ratio of \Teff and radii within the stellar pair. Because  the very hot evolved stars in binaries with an optically brighter companion are elusive at all wavelengths except the UV (Figure \ref{f_SEDbina}), this work offers a unique opportunity to estimate the binary fraction of hot compact stars from an unprecedentedly large sample, two orders of magnitude larger than the ensemble of the detailed studies on which our knowledge of stellar multiplicity for early-type stars is based.
The binary-candidate sample identified in this work is $\sim$4$\times$ larger numerically than the astrometric WD-binaries detected by Gaia (about 3,200 WD-binaries, of which about 110 with a known UV excess)  and comparable to the number of Gaia wide (resolved) WD binaries, about 16,000  (\citet{shahaf24,elbadry24gaia,garbutt24} and references therein). Given the paramount progress enabled by Gaia in this field, it is important to point out that our sample extends the Gaia WD-binary census not simply numerically, but with a higher sensitivity to the sub-population that may be elusive to Gaia.

\begin{figure}[h]
\includegraphics[angle=0,width=9.5cm]{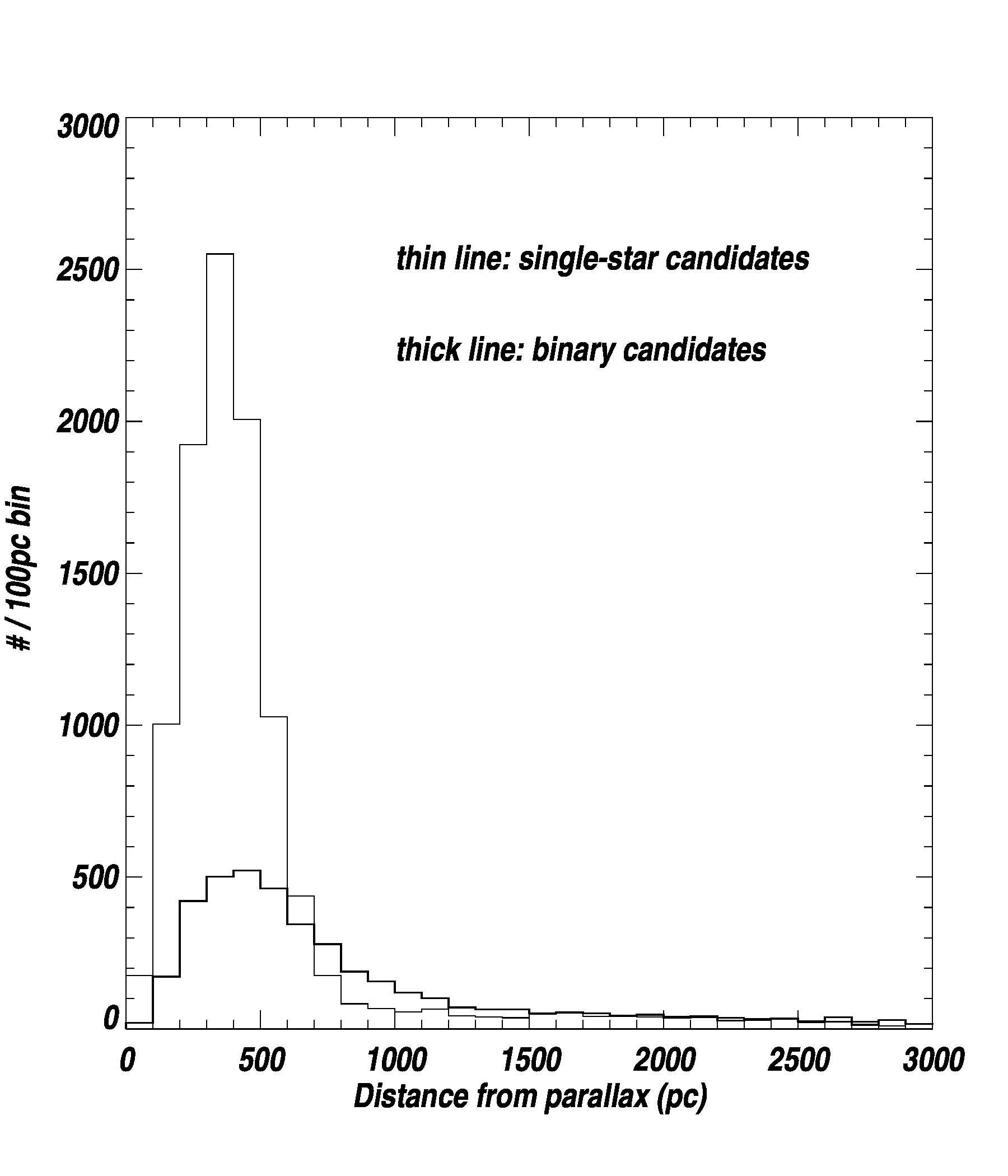}
\caption{Distances derived from Gaia DR3 parallax for the binary- and single-star candidates that have a Gaia match and a  parallax measurement with error $\leq$20\%, shown with a thin and thick line respectively. Actual values in the sample stretch to $\sim$6kpc, with very few data-points beyond the range shown. 
 \label{f_distance} }
\end{figure}

It is interesting to look at the Gaia data among binary- and single-star candidates. Out of the 35,294 sources in the culled analysis sample, 95\% (33,531 sources) have a Gaia match, and 82\% of these (28,791 sources) have also a parallax measurement, of which only 14,490 sources (41\%) with parallax error better than 20\%. In more detail,  the fraction of sources with a good parallax measurement is significantly higher for the single-star candidates (45\%) than for the binary candidates (34\%).   The higher fraction of good parallax measurements among single WDs seems counter-intuitive, because hot compact objects have low optical luminosity (owing to their small radii), and the Gaia limit for detecting nearby WDs is stretched to larger distances when there is a main-sequence or giant late-type companion, which is much brighter than a hot WD at optical wavelengths, hence it is measureable by Gaia farther than the WD. In fact, the distribution of parallax-derived distances (Figure \ref{f_distance}) shows that, among the sources with a good Gaia parallax (error $\leq$20\%) binary candidates are identified at larger distances than single hot-star candidates. The relatively fewer good parallax measurements available among the binaries could then be explained if most systems were unresolved or semi-resolved, making it difficult to solve for parallax, proper motion and orbital motion, and possibly also making the position measurements more uncertain than for a single, perfectly point-like source. We recall that DR3 can resolve nearby sources with angular separation down to 0.18\as in general;  however, the hot WD can be much fainter than the close companion at optical wavelengths (Figure \ref{f_SEDbina}).  Such possibility is  supported also by comparing the fraction of sources with a parallax measurement, regardless of error: 10,751 or 87\% of the 12,404 sources in the binary locus, 20,789 or 91\% of the 22,848 single-star candidates. Among the Gaia matches with a  reported parallax value, as many as 
1,676 (16\%) have a negative parallax value in the binaries sample, versus 5\%  (1,102)  in the single-star sample. Finally, of the 9,075  parallaxes with a positive value for binaries (19,687 for singles), 4213 (34\%),  and 10,261 (46\%) have parallax error $\leq$20\% for binary- and single candidates respectively. 

In sum, among the Gaia counterparts there are relatively fewer good parallax measurements for binary candidates than for sources with single-star-like SED. As mentioned, binarity can offer an explanation for the difficulty of measuring parallaxes, especially if most systems are very close (projected on the sky), as is expected given the distribution of distances and the generally lower quality of parallax measurements among binary candidates.  An alternative explanation for the relatively higher number of Gaia sources with no parallax measurements could be the presence of point-like QSOs or AGNs, whose colors intrude part of the binaries color locus (Figures \ref{f_ccwide} to \ref{f_ccselFN}), as discussed in Section \ref{s_binaccd}. If all the binary candidates lacking a parallax measurement were extra-Galactic objects, the contamination would be 13\%, close to the conservative limit we estimated based on the sources with Simbad identification. If  all the Gaia sources with a negative parallax values were also extra-Galactic objects, the correction to the binary-candidate counts would be 27\%.

We also recall that half of the binary candidates identified by this study are unknown objects. Only 6,030 out   of  the 12,404 binary candidates have a known object match in Simbad (including the 941 QSOs or AGNs or Seyfert1), and  2,651 of these have a spectral type;  also, in several of the known stars the WD companion is previously undetected, whereas only $\sim$20\%  of the single-star candidates are un-known objects: 18,634 (7,148 with a spectral type) out of 22,848 sources have a Simbad match.

In  Section \ref{s_binaccd} we have counted the number of candidate single hot stars and of binaries (hot star plus a cooler companion) using their multi-band color separation, as informed by stellar model colors and by the subsample with a known classification, and have discussed the limits and biases of the relative counts. Some ``single'' sources may be the result of binary merging in prior phases, which is what ultimately we would like to estimate, by comparison with binary fractions of their progenitors at earlier phases. The merging ratios inform theory of binary evolution and stellar population models.  But the sources in the ``single-star'' color locus could also be pairs consisting of two identical (or almost identical) stars, whose \teff s do not differ enough to make the broad-band colors appear like a two-component SED in the wavelength range available,  within typical photometric uncertainties of the present sample (see Figure \ref{f_SEDMSandWD_SDSS}).  For the sub-sample with good Gaia distances, \Rstar and \Lbol can be derived from SED-fitting, once \Teff and \Ebv are derived, as shown in Section \ref{s_sedfit}; however, for a hot WD of a given \teff, the radius \Rstar  may easily vary by a factor of two or more, depending on its mass and \logg , making it impossible to identify ``identical twins'' with apparently single-star SED based on their luminosity.  Therefore, the counts of sources in the ``singles'' locus is an upper limit to the actual number of WDs evolved from initially single stars. 
By consequence, the estimated binary fraction would be a lower limit, if we have reasonably corrected the sources in the binary locus the from QSOs contamination. 
As shown in Figures \ref{f_ccwide} $-$ \ref{f_ccselFN}, a small region of the colors locus of  binaries consisting of a hot WD and a cooler, less evolved star, is also shared by low-redshift QSOs or AGN. 
 Because half of the binary-candidate sources are unknown objects, the estimate of the contamination by extra-Galactic objects is uncertain. Among the known objects (with classification in Simbad), the number of QSOs+AGN+Seyfert are 7.7\% of the binary candidates; if the relative proportions were the same among the unclassified sources, which are about half of the binary candidates, the contamination would be   $\sim$15\%.  Assuming this correction to the binary counts (0.85\% of 12,404), the derived binary fraction is \fbina$\geq$46\%   in the range examined.   The derived binary fraction for our specific sample is much higher than earlier estimates, which reported 18\% to 26\% of WDs to be in binary systems \citep{holberg09,toonen17}, and is higher for example than the wide-binaries fraction derived for polluted WDs and field WDs by \citet{noor24},  of about 10\% or less. 
Future work should characterize the stellar parameters of individual pairs, when possible (examples of preliminary results were given in Figure \ref{f_SEDbina}),  to also explain the differences between the binary fraction derived in our specific sample versus that found in other samples selected in different ways and with different criteria.

  In the count of binaries, companions with spectral type  earlier than $\sim$A cannot be always distinguished from single stars (depending on the ratio of radii, etc.); therefore, some types of binaries would be missed. 
However, because the IMF is skewed towards lower masses, this bias should not be significant because the later-types, lower-mass companions should be much more numerous. Single stars of intermediate and late types are eliminated by the FUV-NUV color cut of the sample, that includes sources hotter than \Teff$\sim$15,000-20,000K, depending on gravity (Figure \ref{f_fuvnuv}). 

 The measured binary fraction is \fbina$\gtrsim$50 or $\gtrsim$46\%, after correcting the binary-candidate counts for a 7.7 or 15\% contamination by   extra-Galactic objects. Among the hottest sources in the sample, with FUV-NUV$\leq$-0.3mag (corresponding to \Teff$\gtrsim$30,000K, depending on gravity), the binary  fraction is higher,   \fbina$\sim$57\%.  In this UV color range, the contamination by extra-Galactic objects is negligible, based on the known samples.  Because  the single-star counts could include  binaries consisting of two similar stars, this fraction is a lower limit.  Compared with the binary fraction of  intermediate-mass unevolved stars reported by \citet{moe2019} (\fbina~ from $>$80\% to 50\% in the mass range 8$-$1\Msun), this result implies that $\lesssim$20\%  of the initial binaries merge, or much less if most of the current WDs' progenitors were closer to the lower mass range of \citet{moe2019}, as is likely. Such implied merging-rate is  
 lower than the result obtained by \citet{sanaetal2012} based on the orbital period distribution of their O-stars sample: they estimate that $\sim$23\% of the very massive stars merge during their lifetime.  However, if the contamination by extra-Galactic objects in our binary counts were underestimated,   \fbina~ would be lower, and the implied merging rate higher.

\section{Summary and Conclusions}
\label{s_sum}

\subsection{Summary of Results, of Catalog Features and Unique Advantages}
\label{s_power}

We have constructed a catalog of hot 71,364 UV GALEX sources with FUV-NUV$\leq$0.1mag and optical SDSS counterparts, deemed point-like at the SDSS resolution, with additional data to facilitate science analysis. The sources are mostly stellar,  with some small contamination by low redshift QSOs or AGNs, and are  mostly hot WDs and hot SDs. We reduced the sample to 35,294 sources with good SDSS photometry for analysis (Section \ref{s_culling}).  From the UV-optical colors in the seven bands FUV, NUV, {\it u, g, r, i, z}, analyzed with model-color grids, we identified in the sample 12,404$\pm$$^{1871}_{1267}$ 
candidate binaries, consisting of a hot WD or SD and a less evolved (in most cases) companion of spectral type later than $\sim$A (the exact limit depending on gravity, and ratio of stellar radii within the pair, e.g. Figure \ref{f_SEDMSandWD_SDSS}), and 22,848$\pm$$^{1267}_{3853}$
 sources with an apparently single-star SED, that could be largely {\it bona fide} single hot stars but could also contain binaries with very similar stellar parameters, or  with an unevolved companion not detectable in the wavelength range of the UV$-$optical SED. 
 
After correcting for an estimated $\sim$7.7$-$15\% contamination of the binary candidates by extra-Galactic QSOs and AGNs, the resulting binary fraction is $\sim$50$-$46\% for the sample (\Teff $\gtrsim$15,000K), and $\sim$57\% for the sources with FUV-NUV$\leq$-0.3mag, i.e. with  \Teff hotter than $\sim$20,000-30,000K (the \Teff limit depends on gravity, see Figure \ref{f_fuvnuv}).  The estimate of \Teff is rather robust,  given that the GALEX FUV-NUV color is essentially reddening-free for Milky-Way type dust with \rv=3.1 (Section \ref{s_intro}). However, the reddening \ebv, and its uncertainty,  affects the estimate of stellar Radius and \lbol, which is obtained by scaling the best-fit model SED to the observed magnitudes, accounting for reddening. The derivation of individual stellar parameters will be the subject of a future work. 

Only a fraction of the sources has a Gaia DR3 identification, and only 34\% of the binary candidates and 45\% of the single-star candidates have a parallax with error better than 20\%, in spite the binaries span larger distances (as estimated from the subsample with good parallax measurements, see Figure \ref{f_distance}). The subsample with a good distance estimate is particularly valuable because, besides \Teff and extinction,  stellar radius and \Lbol can also be derived from the observed SED (Section \ref{s_sedfit}).   Another interesting result is that about half of the entire sample has a known-object counterpart in Simbad; but, in the culled sample that was analyzed here,  the fraction of previously known objects is $\sim$81\% for the single-star candidates, and only 48\% for the candidate binaries; in addition, among these known objects the presence of a hot companion was often unreported.  This large discrepancy is not surprising, because the presence of  a hot WD in a close binary system  with a cooler unevolved companion would be mostly undetected unless FUV data are available, as is evident from the examples in Figure \ref{f_SEDbina}.   

In sum, in our census of binary candidates,  only $\sim$48\%  are
previously known objects  according to the Simbad database,  while  $\sim$ 80\% of the [apparently] single evolved hot-star candidates are known objects.
But among the known objects in our binary sample, the hot-WD companion was often undetected prior to this work.  Among the Gaia DR3 matches to our culled analysis sample of  35,294 UV sources, only forty are flagged as ``NON\_SINGLE\_STAR'', of which thirty-five are selected as binary candidates in this work, and five as ``singles''.  In the less restricted sample of  46,744 sources we only find five more Gaia matches tagged as ``NON\_SINGLE\_STAR', fortyfive in total.  In the analysis sample,  sixtytwo (seventytwo in the larger sample)  have DR3's tag IN\_VARI\_ECLIPSING\_BINARY set to ``True''; fifty-one of the sixty-two are binary candidates according to our selection and eleven are single-WD candidates, confirming that - as previously explained - the ``singles'' locus may contain some types of binaries, while the selection of binary candidates is very robust at least for hot WDs with companions cooler than $\sim$8-10,000K.  The Gaia census of binaries, including those with a hot WD component, is expected to increase significantly by the end of the mission \citep{elbadry24gaia}.  However, the comparison of our results with the current identification of binaries  from the DR3 Gaia pipeline  highlights the unique leverage offered by our sample with UV$-$optical wavelength coverage. In addition to the DR3 classification, several  recent works used various combinations of criteria to identify different types of binaries from the Gaia data. 
\citet{shahaf24} identified  $\sim$3,200 astrometric WD binares, about one fourth of our binary-candidate sample size,  and about 16,000 wide WD binaries have been reported (\citet{elbadry24gaia} and references therein). More relevant than the numerical differences, which are difficult to interpret because each method uses specific criteria that select subsamples with different characteristics,  is the complementarity of the selections, in particular the leverage offered by the UV catalog to identify WD-binaries in a parameter regime (very hot, or very tight orbit, or low-mass)  that may elude Gaia's selections, contributing a different piece of the puzzle towards a complete picture (see also \citet{elbadry24gaia} for a discussion on Gaia results). 

The manyfold increase of the census of binaries with a hot compact object is relevant because a
comparison between the binary fraction of hot evolved objects, estimated for the first time from a large statistical sample,
with the binary fraction of their progenitors, the massive and intermediate-mass main sequence stars, translates into the percentage of the initial binaries that merge before one component reaches the post-AGB phase. 
Intermediate-mass stars are the major providers of C, N and other abundant elements, including elements supporting life as we know it. Statistically significant, unbiased samples of their evolved descendants could help clarify their evolutionary path (in particular the yield of chemical elements from the so called ``third dredge-up''), ultimately relevant for understanding  the chemical evolution of the universe,  as well as to inform star formation and stellar evolution theories, which underpin our interpretation of the integrated colors of galaxies at cosmological distances, where  stellar populations cannot be resolved into their individual stellar constituents. 
Studies of stellar binarity are being extended to the Magellanic clouds to explore effects of metallicity (e.g., \citet{dordapatrick21,patrick19,patrick20,patrick22} with samples of cool supergiants having a hotter ($\sim$B-type) companion, where a binary fraction of $\sim$15$\pm$4\% (SMC) and 14$\pm$5\% (LMC) was found by  \citet{dordapatrick21}.  The catalog presented here  allows the investigation to be extended to the hot post-AGB population, so far only accessible in the Milky Way, and in particular to extend the census of hot-WD in binary systems in the parameters ranges that make them elusive to Gaia detection methods, potentially filling gaps in poorly explored regimes.

Only a fraction of our UV-optical source catalog has a Gaia DR3 match, in spite it was extracted from the GALEX shallowest sky survey (AIS) with rather bright  UV magnitude limits  (see e.g., \citet{bia09,bia14uvsky,bia17guvcat}). In the wider catalog of 71,364 sources, of the 59,249 input UV  
sources that have at least one Gaia match, i.e. the primary matches that are not ``null'' (null matches have GAIA\_ID = '-888' in the master catalog), 53,331 have only one match within 3\as, and only 2,927 have additional matches. Such figures imply that in the vast majority of cases there is no ambiguity on  the actual Gaia counterpart of the GALEX source.
Out of the 35,294 sources in the culled analysis sample, 95\% (33,531 sources) have a Gaia match, and 82\% of these (28,791 sources) have also a parallax measurement, 14,490 of which (41\%) with error better than 20\%. In more detail,  the fraction of sources with a good parallax measurement is higher for the single-star candidates (45\%) than for the binary candidates (34\%) (Section \ref{s_binaf}).  Because most of the sources are hot WDs, these numbers show the power of the UV source catalog to  identify such very hot, optically-faint stellar objects, and how elusive they are even in the best available optical surveys todate.

The unique leverage of the UV catalog to identify hot WDs is also evident from the statistics of known objects, resulting from the match with the  Simbad database (Section \ref{s_simbad}). The majority of the known objects in our catalog   are classified as WD*\_Candidate (14,830 matches) and similar stellar classes, according to the Simbad database, highlighting the advantage of the UV-color selection with respect to other wavelengths.  On the other hand, only about half of our catalog sources are known objects (27,440, or 58.7\% in the conservatively culled analysis sample), and only 11,315 have a reported spectral type, highlighting again the power of the presented UV catalog to extend the known samples of elusive hot compact stars, with significant purity.  As reported in Section \ref{s_binaccd}, there is a remarkable  difference between the fraction of known objects in the single- (80\%) or binary- (49\%) candidates.  The UV colors are particularly relevant to identify hot WDs in binary systems with a cooler, larger, optically brighter companion, whose flux  makes the hot WD contribution negligible, hence its presence undetectable, at optical and IR wavelengths (examples in Figure \ref{f_SEDbina}).

 The initial selection in this work was made for sources with FUV-NUV$\leq$0.1mag, and with a SDSS match. In the restricted 46,744 analysis sample (regardless of SDSS data quality), 20,782 (44\%) 
have FUV-NUV $\leq$-0.2~mag, and 28,279   (60\%) 
have FUV-NUV $\leq$-0.3~mag. Therefore, the present sample contains an unprecedented high number of hot WDs (see Figure \ref{f_fuvnuv}) with respect to known samples.

\subsection{Possible Applications and Follow-up Work}
\label{s_futurework}

Because we do not know the orbital  separation of the  binary candidates (all pairs are unresolved in GALEX and SDSS imaging), we cannot identify which of the binaries found in this work have likely exchanged mass, and which ones are separated enough that they can be reasonably assumed to both have evolved as a single star. While recent efforts to understand binary evolution concentrate on the effects of mass exchange (e.g., \citet{demink2013,demink2014,sen2022}), which affects the majority of massive stars, the binaries that have {\it not} exchanged mass can offer insight into the initial-final mass relation, that maps the WD mass to the initial mass of its progenitor.  If no mass transfer has taken place, the mass of the cooler, unevolved companion is a lower limit to the mass of the initially more massive WD progenitor; and when the evolutionary age can be inferred from the cooler companion, the mass of the WD progenitor can be further constrained, assuming the pair was born coeval (physical binary, not the result of a capture). Comparison of the evolutionary time on the cooling track (from the current WD location in post-AGB HR-diagrams) with the total age anchored to the less evolved companion, yields the presumed pre-AGB lifetime of the WD progenitor, and constrains its mass.
We are pursuing the identification of binaries that evolved with no mass exchange  in two HST programs (Bianchi's HST-GO-14119 and HST-GO-15832), with HST high resolution imaging of over 100 candidates, selected from the broad sample presented here. We have identified about two dozens such cases, currently being analyzed (\citet{bianchiSNAP}).    

Besides the ongoing follow-up with HST of the selected subsamples, possible future work will explore  a machine  learning (ML) approach to better separate single and binary candidates, and to better estimate the limits and biases, based on the model colors and the SED of known objects in the catalog.   Other follow-up work such as radial-velocity monitoring may be of interest to others, to characterize our UV-selected sample with respect to samples identified by other methods.

Finally, the presented sample, and the derived binary fraction among post-AGB objects, can contribute to clarifying the formation of planetary nebulae (PNe). In particular, two aspects are long debated and still need firm observational constraints. The first concerns the formation of PNe from low-mass stars (mass $\leq$2.3$\pm$0.3\Msun):  the expected stellar mass loss is insufficient to shed enough mass to reach the PN central star (CSPN) stage, unless it is enhanced by mass transfer or common envelope, related to binarity of the CSPN \citep{MoeDeMarco2011}. The second addresses the statistical distribution of axy-simmetric PNe: to produce bi-polar PN morphology, the formation of an equatorial disk is postulated, when mass loss in the previous phases is enhanced by rotation. The equatorial overdensity would allow the PN shell to expand more in the polar directions when the circumstellar material is plowed outwards by the CSPN radiation-pressure-driven supersonic wind in the post-AGB phase  (e.g., \citet{herald11,keller11,bia12} and references therein).  Again, the equatorial overdensity of the AGB-expelled circumstellar material  is believed to be insufficient to account for the observed distribution of PNe geometries; but the preferentially-equatorial  mass loss can be easily enhanced in very close binary CSPNe, possibly  contributing to the formation of bipolar PN shells (e.g., \citet{gomezmunoz23,aliCSPN} and references therein).

\subsection{Limitations and Caveats of the Catalog}
\label{s_caveat}

\subsubsection{There Are Many More GALEX Sources!}
\label{s_more} 

 The sample presented in this work represents a homogeneous, but not complete, subset of the available GALEX UV sources. The sample was extracted from the GALEX AIS survey, which has the shallowest depth but the widest area coverage, as shown in Figure 1 of \citet{bia14uvsky}, and in particular only from the AIS fields that have both detectors (FUV and NUV) exposed. We used as a starting point $GUVcat\_AIS$ \citep{bia17guvcat} because it is a homogeneous catalog of unique GALEX sources  in which some sources that have incorrect measurements in the official GALEX GR6+7 online release have been corrected, and repeated observations of the same source have been distilled to a unique-source list, suitable for matching with databases at other wavelengths and for statistical studies. 
 The sample was further restricted to the availablity of SDSS coverage, about one fourth of the $GUVcat\_AIS$ sources.  We have used the $GUVcat\_AIS$ matched with SDSS ($GUVmatch\_AISxSDSSdr14$, \citet{guvmatch}), because our goal of distinguishing single hot WDs $vs$ binaries requires SED wavelength coverage spanning from FUV to near-IR wavelengths.  The SDSS match also restricts the survey area covered by the extracted sample to the overlap of the two surveys, given the limited SDSS coverage (shown by \citet{bia14uvsky}: their Figure1 bottom; the overlap area is $\approx$11,000 square degrees, from \citet{areacat}).  $GUVmatch\_AISxSDSSdr14$  contains tags to identify multiple matches, useful to clean the extracted sample, and the five optical bands provided by SDSS allow us to select binaries in interesting, as well as poorly explored, regimes, and to estimate their parameters. 
  278,375 sources with FUV-NUV$\leq$0.1mag are extracted from the matched  $GUVmatch\_AISxSDSSdr14$ database ($\sim$22.2 million) and further restricted to 71,394 point-like sources.  Many more UV sources exist in the GALEX database, with FUV-NUV$>$0.1mag, and many more (at least 10$\times$) with NUV-only measurements:  $\approx$83~million sources just from the AIS survey \citep{bia17guvcat}.  

\subsubsection{Variable Sources}
\label{s_var}
An obvious warning concerns variable sources. The GALEX, SDSS, and Gaia data have been taken at different epochs, up to two decades apart. For stellar sources,  such as, e.g.,  cataclismic variables (CVs) and other interacting or eclipsing binaries, and for any type of variable sources (including QSOs),  colors from  combined GALEX+SDSS SED or GALEX+Gaia photometry would be misleading if the source varied between the times of the respective observations.   As described in Section \ref{s_gaia}, we have also linked the Gaia DR3 main catalog with the Gaia variability information, which is included in the master catalog. Out of the 35,294 sources in the culled sample analyzed in Section \ref{s_analysis}, only 1,763 sources do not have a Gaia match (tag GAIA\_ID = -888). Among the 33,531 sources with a Gaia counterpart, 2,104 have the tag PHOT\_VARIABLE\_FLAG set to ``VARIABLE''.  These sources are interesting for other purposes,  but their SED could have biased UV$-$optical colors in the matched catalog, where only average magnitudes from all existing observations are recorded. Gaia DR3 tag IN\_VARI\_CLASSIFICATION\_RESULT is set to ``True'' for 1,971 of them; these include 62 variables flagged as IN\_VARI\_ECLIPSING\_BINARY, 24 as IN\_VARI\_SHORT\_TIMESCALE,  39 as IN\_VARI\_ROTATION\_MODULATION, 3 as  IN\_VARI\_RRLYRAE, 3 as IN\_VARI\_MS\_OSCILLATOR. Interestingly, the highest number of Gaia counterparts with reported variability, 1,028,  are flagged as IN\_VARI\_AGN. This number is very close to the 954  known extra-Galactic objects found from the Simbad classification in the binaries color-locus. Indeed, most of the sources with Gaia counterpart with flag IN\_VARI\_AGN set are those classified in Simbad as ``QSO'' or ``Seyfert\_1''   (very few as ``Blazar'' or ``BLLAC'' or ``AGN''), and only 228 do not have a Simbad object classification.  This comparison also suggests that the contamination of the binary candidates by extra-Galactic objects may be closer to the 8\% estimated from the known objects in Simbad than to twice as much, which we have conservatively used to correct the sample of candidate variables.

\clearpage 

\section{Appendix A.  Match of the point-source sample with individual GALEX observations.}
\label{s_appendixA} 

As mentioned in Section \ref{s_gaia}, we matched the selected  catalog of point-sources with all GALEX data at visit level (i.e., measurements from individual observations), cross-matching the 
$GUVcat\_AISxSDSS\_HSpoint$  catalog with table $visitphotoobjall$ in the Casjobs database\footnote{in \url{mastweb.stsci.edu/mcasjobs}, Context: GALEXGR6\_plus7, \url{https://mastweb.stsci.edu/mcasjobs/mydbcontent.aspx?ObjName=visitphotoobjall\&ObjType=TABLE\&context=GALEX\_GR6Plus7\&type=normal} the $visitphotoobjall$ table has 572,968,390 visit-level source measurements}. 
A 3\as match radius was used.  When multiple visits were found for the same source, we selected the ``best'' visit, defined as the observation with the longest exposure time with both detectors exposed, excluding - when possible - visits where the source was on the edge of the field, which suffers from some distortion, and then  used  the RA,DEC,Epoch of the GALEX best visit for the Gaia cross-matching.   The 
full match output of the individual GALEX visits is made available, given that a number of sources have multiple observations, and these can be useful for different purposes, including a limited serendipitous variability search between repeated observations, or for examining different images of specific sources of interest.

The match returned 352,642 GALEX sources measured in individual visits, within 3\as of the 71,364 input point sources, indicating that multiple observations exist for a number of sources; these repeated visits allow a serendipitous variability search.  For 25 input sources no counterpart was found within 3\as in the visit-level source catalog. These sources are kept in the output for completeness. 
 The input catalog was extracted from sources measured mostly in coadded images of available observations (but see \citet{bia17guvcat} about issues with some coadded images), which may result in a detection in the coadded data even for sources undetected in individual images; also, for a source in a complex field or for an extended object, the pipeline may center the source in the coadded image at a position RA,DEC which differs by more than 3\as (our match radius) from potential counterparts identified in the individual visits. All such cases probably are not bona-fide point-like sources, or well-resolved sources; therefore, we eventually eliminate them from the analysis sample (Section \ref{s_culling}) although they are kept in the general point-like catalog for completeness, to indicate that GALEX images in those positions exist and can be used for custom-photometry if the source is of interest.

 For most of the GALEX visits both FUV and NUV detectors were on, but there are cases in which one of the two was not exposed (mostly FUV, that stopped working in 2009 but was also off occasionally in the earlier part of the mission). We recall, as pointed out by \citet{bia11a,bia11b,bia17guvcat} that a non-detection of a source in one of the bands may mean that that detector was not exposed, or that both detectors were exposed but the source was too faint to be detected in one of the two. Such important distinction cannot be found from the GALEX standard database catalogs (both non-exposure and non-detection have magnitude~=~-999) unless the exposure times are extracted; we provide exposure times in all our catalogs.  On average, only about 10\% of the NUV sources are also detected in FUV, for comparable exposure times, because the stellar and galaxy populations are skewed towards cooler sources, and ``hot'' (i.e., FUV-bright) sources are more rare \citep{bia14uvsky,bia17guvcat}. However,  there are also cases where the pipeline detected a source in the FUV image and not a counterpart in NUV; such cases mostly occur because NUV sources, being more numerous, are also more crowded, and in dense fields such as, e.g., in stellar clusters, they can be so crowded that the pipeline merges several nearby sources into one extended source in the NUV image, while it resolves them in the FUV image. This is illustrated in Figure 5 of \citet{bia17guvcat}. 
Of the 352,617 
non-null visit matches, 117,878 have no FUV measurements and 462 have no NUV measurement. 

 Relevant parameters of the GALEX ``best-visit'' are also included in the master catalog for point sources, with  
 tags: {\tiny{ GALEX\_BESTVIS\_OBSDATE, GALEX\_BESTVIS\_NUVTIME, GALEX\_BESTVIS\_FUVTIME, GALEX\_BESTVIS\_EPOCHDECIMAL, GALEX\_BESTVIS\_FUV\_WEIGHT, GALEX\_BESTVIS\_NUV\_WEIGHT, GALEX\_BESTVIS\_FOV\_RADIUS, GALEX\_BESTVIS\_RA, GALEX\_BESTVIS\_DEC, GALEX\_BESTVIS\_FUVMAG, GALEX\_BESTVIS\_FUVMAGERR, GALEX\_BESTVIS\_NUVMAG, GALEX\_BESTVIS\_NUVMAGERR, GALEX\_NVIS\_FUV, GALEX\_NVIS\_NUV, GALEX\_MIN\_FUVMAG, GALEX\_MIN\_FUVMAGERR, GALEX\_MAX\_FUVMAG, GALEX\_MAX\_FUVMAGERR, GALEX\_MIN\_NUVMAG, GALEX\_MIN\_NUVMAGERR, GALEX\_MAX\_NUVMAG, and GALEX\_MAX\_NUVMAGERR}}. The tags are described in Table \ref{t_tagspoint}.
 Tags starting with ``GALEX\_BESTVIS'' are values imported from $visitphotoobjall$ and $visitphotoextract$.  We also constructed tags GALEX\_NVIS\_FUV and GALEX\_NVIS\_NUV that give the number of images in which the source is found that have the FUV and NUV detector exposed, respectively. Note that in some cases the source is not detected, because the exposure is too short and the source may fall below detection threshold.  In this case, the minimum value among all measured magnitudes 
(GALEX\_MIN\_FUVMAG or GALEX\_MIN\_NUVMAG) is -999.  We included in the report of minimum and maximum values also the non-detections, because they can be informative of variability. Instead, to select the ``best visit'', we examined only the visits where both the exposure time and the measured magnitude are above zero. If such cases exist, the ``best visit'' is chosen as the one with the longest (NUV) exposure time; if no visit exists with both detectors exposed and the source detected in both bands, then the ``best visit'' is chosen among all NUV exposures regardless of FUV. The measurements from this visit are propagated in the master catalog in the tags listed above. 

  One extreme example of a faint source can illustrate the -sometimes complicated- procedure to distill the visit-level information into the master catalog. 
Source GALEX\_ID = 6376686083780907764 has five visits with both FUV and NUV detectors exposed.  The FUV effective exposure times are FUV\_WEIGHT= 59.8594,      60.2578,      78.9844,      56.8984,  and     65.2344~sec, and the respective  FUV\_MAG are = -999.000,      21.9650,      21.4533,     -999.000, and     -999.000~ABmag. That is, in three of the five FUV images this source does not have a significant detection.  
For NUV, the exposure times  are NUV\_WEIGHT=78.1406,     78.1719,      68.7969,      71.5938, and   67.3125~sec, and in the second visit the source is not  detected: NUV\_MAG= 21.4124,     -999.000,      21.7302,      21.3130,      21.5072~ABmag. So, there is only one visit with both detectors exposed {\it and} the source detected in both; therefore, this is chosen as the ``best visit'', resulting in tags:  GALEX\_BESTVIS\_FUVMAG=21.4533~ABmag and
 GALEX\_BESTVIS\_NUVMAG=21.7302~ABmag, with all the other ``best visit'' related parameters (error, position, epoch, exposure time).   However, the resulting minimum and maximum value among all existing visits with non null exposure time are  GALEX\_MIN\_FUVMAG=-999. and  GALEX\_MAX\_FUVMAG=21.9650~ABmag, GALEX\_MIN\_NUVMAG=-999. and  GALEX\_MAX\_NUVMAG=21.7302~ABmag. Among the five repeated visits,  SEP=0.49210\as 
 for the only visit where the source was detected in both FUV and NUV, and SEP=-999.\as for the others.  We expounded here on the methodology and the criteria for clarity, for a corrrect interpretation of the TAGS provided in our catalog, and for future use also by others who may want to serendipitously search for variability among the repeated observations of sources in the GALEX databases.

 We point out here another caveat, not previoulsy reported, because it may help users of the standard GALEX database, in particular 
 when using misleading information from  Casjobs table    $visitphotoobjall$. In this table, the tag ``BAND'' indicates which detectors were exposed during the visit; 
namely, BAND=1 indicates that NUV was exposed, BAND=2 that  FUV was exposed, BAND=3 that both detectors were exposed. However, this information is not consistent with the exposure times FUV\_weight and NUV\_weight. For example, of the 71,364 input point sources, after excluding the 25 that do not have a visit match within 3\as as mentioned, 8,075  have only one match within 3\as in $visitphotoobjall$. 
Of these 8,075 individual visits, 7,989 have BAND=3 in the visit-level catalog (and all have FUV and NUV exposures $>$0.sec, as expected), 75 have BAND=1, and all have NUV exposure $>$0.sec but 61 of these also have FUV exposure $>$0.sec, i.e. both detectors were exposed but the observation was not recorded with BAND=3.  Finally, 11 of these  visits are catalogued with BAND=2, but both detectors were exposed in all 11 cases.  This example only indicates that the tag ``BAND'' (in table $visitphotoobjall$ from the GALEX pipeline) cannot be used to select individual observations according to FUV-NUV color availability.  We recall that our specific sample was selected by a cut in FUV-NUV color; therefore, all sources that have only one visit should have had both detectors exposed in that visit, logically. In the case of the 14 sources with  BAND=1 and no FUV exposure within 3\as, probably the source was extracted from a coadd of different visits and merged with a separation larger than 3\as between FUV and NUV detections (we recall that source detection is performed by the pipeline separately in the FUV and in the NUV image, then the FUV and NUV detections are merged; see discussion about the ``SEP'' tag in Section \ref{s_culling}). Here we have been referring to the BAND tag as recorded in the individual visits; the tag ``BAND'' in the master catalog, however, is propagated from the merged catalog $GUVmatch$, which was built on the coadded database ($GUVcat$ \citep{bia17guvcat}, mostly from $photoObjall$), and because we selected sources with a FUV-NUV cut, i.e. that have necessarily a measurement in both detectors, in the master catalog BAND is always =3.

 In the point-source master catalog 
 25 sources have GALEX\_NVIS\_NUV=0,~ 8,166 have  GALEX\_NVIS\_NUV=1, 15,102 have GALEX\_NVIS\_FUV=1 (there are repeated observations in NUV-only more than there are with both detectors),  8,061 sources have both GALEX\_NVIS\_NUV and GALEX\_NVIS\_FUV=1, and 14 have  GALEX\_NVIS\_NUV=1 and GALEX\_NVIS\_FUV=0 (8 of which with SEP $\leq$ 3.0\as, where SEP is the separation between the NUV-detection and FUV-detection position of the merged sources, not in the individual visits).

The full match with the GALEX observations at visit-level can be found at: \\
\url{http://dolomiti.pha.jhu.edu/hotwd/GUVHScat/GUVcat\_AISxSDSS\_HSpointXvisits\_3arcsec.fits}\\
\url{http://dolomiti.pha.jhu.edu/hotwd/GUVHScat/GUVcat\_AISxSDSS\_HSpointXvisits\_3arcsec.csv}\\

The catalog contains 386 tags (columns), the first six: GALEX\_ID, GALEX\_RA, GALEX\_DEC, GALEXID, GALEXRA, and GALEXDEC are the input catalog source ID and coordinates (repeated twice for database convenience), the seventh, OBJID, is the matched-source ID from $visitphotobjall$ (the visit-level matched source),  followed by tags taken from Casjobs table {\it visitphotobjall}, except for the last tags  GALEX\_ID\_VIS, DISTARCMIN, GALEX\_RA\_VIS, GALEX\_DEC\_VIS, where  DISTARCMIN is the separation between the input and the match, VPE\_NOBS\_DAT, VPE\_NTIME\_OB, VPE\_FOBS\_DAT, VPE\_FTIME\_OB, VPE\_ECLIPSE, and VPE\_LEG are useful parameters (date and time of the observation, ``ECLIPSE'' = a unique GALEX identifier for each visit) from Casjobs table $visitphotoextract$, that we linked to the $visitphotoobjall$ matches in order to obtain the epoch of observation, and finally EPOCHNUV\_DECIMAL and EPOCHFUV\_DECIMAL were constructed from the date and time of each observation, to give the epoch in year.decimal, which is convenient for applying the proper motion correction to Gaia (and other) matches in order to register them to the epoch of the GALEX visit.

\section{Appendix B. Available Data Products.}
\label{s_cats}

 Here we list the catalogs resulting from this project and publicly released. They are available from the author's web site at 
\url{http://dolomiti.pha.jhu.edu/uvsky/GUVcatHS/}.   All our data products are also available at MAST as a High Level Science Product via \dataset[DOI: 10.17909/w9k5-tm92]{https://doi.org/10.17909/w9k5-tm92}  
(url:  
\url{https://archive.stsci.edu/hlsp/guvcat-hotstars/}), 
and CDS Vizier.

~\\
{\bf GUVcat\_AISxSDSS\_HSpoint} (71,364 sources)\\
Point-like sources, selected from SDSS tag $type$='STAR' from 278,375 sources in $GUVmatch\_AISxSDSS$ \citep{guvmatch} with FUV-NUV$\leq$0.1mag; no other culling.  This is the ``master catalog'' used as a 
starting point in the present work; it contains, in addition to the tags from the initial $GUVmatch\_AISxSDSS$, additional information to facilitate the science analysis. Specifically, 
columns 223-231 contain distilled information from the Simbad match (Section \ref{s_simbad}), columns 233-257 contain some parameters from the GALEX best visit (individual observation, see Appendix~A) among all the visits from which the source was extracted,  and columns 258-314 contain information distilled from the Gaia DR3 match (Section \ref{s_gaia}) including serendipitous detection of variability in the Gaia DR3 database, and tags constructed to track multiple matches. The columns are described in Table \ref{t_tagspoint}. See section \ref{s_sample} for other useful tags, such as ``INLARGEOBJ''; sources in the footprint of stellar clusters are excluded from the analysis in this work, but retained in the master catalog, and may be of interest for a variety of purposes but require careful checking of photometric quality. 
The catalog can be downloaded at:\\
{\footnotesize 
\url{http://dolomiti.pha.jhu.edu/uvsky/GUVcatHS/GUVcat\_AISxSDSS\_HSpoint.fits}\\
\url{http://dolomiti.pha.jhu.edu/uvsky/GUVcatHS/GUVcat\_AISxSDSS\_HSpoint.csv}\\
 }

~\\
{\bf GUVcat\_AISxSDSS\_HSpointXGaiaDR3}\\
The complete Gaia DR3 match results to the 71,364 
$GUVcat\_AISxSDSS\_HSpoint$ point-like sources. See Section \ref{s_gaia}. It contains 59,249 Gaia counterparts to 56,258 GALEX input sources;  the 10,399 input sources that had no match within 15\as, and the 4,707 that have a match within 15\as but not within 3\as, are excluded. 
Columns 1-222 are from the $GUVcat\_AISxSDSS\_HSpoint$ 
catalog. The subsequent columns contain all the tags from the Gaia main  catalog ({\it source} table in DR3), followed by the tags from the Gaia DR3 {\it vari\_summary} table. 
These files are the full match output, where  all Gaia multiple matches within the match radius are retained, as defined in Section \ref{s_gaia}; in the master catalog we distilled only the Gaia tags of the ``primary'' match, i.e. the only match if MMRANK\_GAIA = 0, or the closest among multiple matches (MMRANK\_GAIA =1), in order to keep a list of unique sources.   Therefore, if multiple matches are found and should be examined, they can be found in this catalog.  For sources with significant proper motion in Gaia DR3, the catalog also contains the Gaia source position ``rewinded'' to the time of the GALEX visit (Section \ref{s_gaia}).  Table \ref{t_gaiatags} explains the columns. 
The full Gaia-match output is available at :\\
{\scriptsize
\url{http://dolomiti.pha.jhu.edu/uvsky/GUVcatHS/GUVcat\_AISxSDSS\_HSpointXGaia3arcsecREWINDED\_noNULLS.fits} \\
\url{http://dolomiti.pha.jhu.edu/uvsky/GUVcatHS/GUVcat\_AISxSDSS\_HSpointXGaia3arcsecREWINDED\_noNULLS.csv} \\
}

~\\
{\bf GUVcat\_AISxSDSS\_HSpointXsimbad\_[5 and 10]arcsec} \\
Full results of the match of $GUVcat\_AISxSDSS\_HSpoint$ 
with the Simbad database (Section \ref{s_simbad}), with match radius of  5\as and 10\as. All the  Simbad columns of the resulting matched sources are appended after columns  1-222 of Table \ref{t_tagspoint}.  The relevant parameters of the nearest Simbad match, and the number of Simbad matches within 5\as and 10\as, are also included in the master catalog of point-like sources, see Table \ref{t_tagspoint} - columns 265-271.   Values are set to ``=='' if there is no match, and to ``-888.'' if no value of a given tag is given for the match. \\  
The catalogs (original outputs from Vizier)\footnote{we report that during this work, the first match obtained with Vizier X-match tool gave incorrect results, i.e, it returned matches with a reported distance smaller than the chosen match radius, but the actual  separation between the coordinates of the input source and of the returned Simbad match were very different than the separation returned by Vizier, and in many case much larger than the match radius; we had incurred in the same problem before, and reported it to the CDS team, who fixed the synchronization with Simbad. We have repeated the match after a few months and checked that these files give correct distances between input source and match} can be downloaded at:\\
{\footnotesize 
\url{http://dolomiti.pha.jhu.edu/uvsky/GUVcatHS/catHSpointXsimbad\_5arcsec\_1681832401503A.csv}\\
\url{http://dolomiti.pha.jhu.edu/uvsky/GUVcatHS/catHSpointXsimbad\_10arcsec\_1681832559998A.csv}\\
}

~\\
{\bf GUVcat\_AISxSDSS\_HSpointXvisits} \\
Match of the master catalog with GALEX database table {\it visitphotoobjall}, to link each source to all its existing GALEX observations (when a match at visit-level is found within 3\as of the master catalog position), as described in Appendix A. The full results are given in this catalog. Some relevant parameters, such as the number of visits found in FUV and NUV, and the best-visit parameters, are also distilled in the master catalog. The file below is the full output of all visits returned with sources within 3\as of the input list: 352,641 rows.   Description for all tags from the {\it visitphotoobjall} table are found in the MAST Casjobs Context ``GALEX GR6plus7'' \footnote{\tiny \url{https://mastweb.stsci.edu/mcasjobs/mydbcontent.aspx?ObjName=visitphotoobjall\&ObjType=TABLE\&context=GALEX\_GR6Plus7\&type=normal} }. Values returned as ``null'' are set to -888.  The catalogs can be downloaded at:  \\
{\scriptsize
\url{http://dolomiti.pha.jhu.edu/uvsky/GUVcatHS/GUVcat\_AISxSDSS\_HSpointXvisits\_3arcsec.fits}\\
\url{http://dolomiti.pha.jhu.edu/uvsky/GUVcatHS/GUVcat\_AISxSDSS\_HSpointXvisits\_3arcsec.csv}\\
}

~\\
{\bf GUVcat\_AISxSDSS\_HSCulled} \\
The analysis sample of 35,294 sources after removing sources with SDSS saturation, possible inconsistencies between the source measurement in the GALEX FUV and NUV images, and trimming to SDSS photometric errors  $\leq$0.2mag in {\it u, g, r, i} (Section \ref{s_culling}). 
This subset of $GUVcat\_AISxSDSS\_HSpoint$, 
as culled for quality in Section \ref{s_culling}, is used for SED analysis in  Section \ref{s_analysis}. It includes all the same tags as the master catalog, described in Table \ref{t_tagspoint}, plus an additional tag ``COLOR\_LOCUS'' that flags whether the source is in the binary-candidate (COLOR\_LOCUS=B) or  single-candidate (COLOR\_LOCUS=S)  color locus defined in Figure \ref{f_ccsel} and described in Section \ref{s_binaccd}.  COLOR\_LOCUS=[B] and [S] indicate sources that are not included in the respective color contours according to their nominal photometry values, but are included when color errors are applied. There are  11,123 sources with COLOR\_LOCUS=B and 1,281 with  COLOR\_LOCUS=''B [S]'' (for a total of 12,404 binary candidates, 1,281 of which overlap with the single-star locus when 1-$\sigma$ color errors are added); 20966 sources have COLOR\_LOCUS=S and  1882 have COLOR\_LOCUS=[B] S (for the total 22,848 single-star candidates from their nominal colors, of which 1,882 falling in the binary-locus when color errors are applied); there are no sources classified as '[B] [S]', and 38 are unclassified.
The catalog can be downloaded at:\\
\url{http://dolomiti.pha.jhu.edu/uvsky/GUVcatHS/GUVcat\_AISxSDSS\_HSculled.csv}\\
\url{http://dolomiti.pha.jhu.edu/uvsky/GUVcatHS/GUVcat\_AISxSDSS\_HSculled.fits}\\

\clearpage

\begin{deluxetable}{lll}
\tabletypesize{\tiny}
\tablecaption{Description of Columns of the Master Catalog of Point Sources $GUVcat\_AISxSDSS\_HSpoint$\tablenotemark{*}  \label{t_tagspoint} }
\tablehead{
\colhead{Col} & \colhead{Column Name} & \colhead{Description} }
\startdata
\tabletypesize{\tiny}
 & & \\
\multicolumn{3}{c}{\bf columns 1-95:  GALEX tags from $GUVmatch$ \citep{guvmatch}}\\ 
1 & GALEX\_ID  & Source identifier from {\it GUVcat} \\
2 & PHOTOEXTRACTID & parent image from which the source was extracted \\ 
3 & MPSTYPE & GALEX survey (AIS, MIS, etc ).\\
4 & AVASPRA & R.A. of the center of the field in which the object was measured \\ 
5 & AVASPDEC & Dec. of the center of the field in which the object was measured\\
6 & FEXPTIME & FUV exposure time (sec) \\
7 & NEXPTIME& NUV exposure time (sec)\\
8 & GALEX\_RA &  J2000 right ascension of the GALEX source from {\it GUVcat}\\ 
9 & GALEX\_DEC & J2000 declination  of the GALEX source  from {\it GUVcat}\\  
10 & GALEX\_GLON  & Galactic longitude of the GALEX source  from {\it GUVcat}\\
11 & GALEX\_GLAT & Galactic latitude of the GALEX source  from {\it GUVcat}\\
12 & TILENUM &  GALEX ''tile'' number (pointing field)\\
13 & IMG & image number (\# exposure for visit) \\
14 & SUBVISIT  & subvisit number if exposure was divided \\
15 &  FOV\_RADIUS  & distance of the source from the center of the field in degrees\\
16 & GALEX\_TYPE  & observation type (0=single, 1=multi)\\
17 & BAND & 1= NUV, 2= FUV, 3=both \footnote{see Appendix A for caveats}\\
18 &  E\_BV  & Galactic reddening from Schlegel et al. (1998) maps \\
19 & ISTHERESPECTRUM  &=1 if there is a spectrum, =0 if not \\
20 & CHKOBJ\_TYPE  & astrometry check type \\
21 & FUV\_MAG  & FUV calibrated magnitude (ABmag system)\\
22 & FUV\_MAGERR  & error of FUV calibrated magnitude\\
23 & NUV\_MAG  & NUV calibrated magnitude (ABmag system)\\
24 & NUV\_MAGERR  & error of NUV calibrated magnitude (ABmag system)\\
25 & FUV\_MAG\_AUTO  & FUV Kron-like elliptical aperture magnitude\\
26  & FUV\_MAGERR\_AUTO & error on FUV AUTO mag\\
27 &  NUV\_MAG\_AUTO  & NUV  Kron-like elliptical aperture magnitude\\
28 & NUV\_MAGERR\_AUTO  & error on NUV  Kron-like elliptical aperture magnitude\\
29 & FUV\_MAG\_APER\_4  & FUV aperture magnitude (8pxl)\\
30 & FUV\_MAGERR\_APER\_4  & error on FUV aperture magnitude (8pxl) \\
31 & NUV\_MAG\_APER\_4  & NUV  aperture magnitude (8pxl)\\
32 & NUV\_MAGERR\_APER\_4  & error on NUV  aperture magnitude (8pxl)\\
33 & FUV\_MAG\_APER\_6  & FUV aperture magnitude (17pxl)\\
34 & FUV\_MAGERR\_APER\_6 & error on FUV aperture magnitude (17pxl)\\
35 &  NUV\_MAG\_APER\_6  & NUV aperture magnitude (17pxl)\\
36 & NUV\_MAGERR\_APER\_6 & error on NUV aperture magnitude (17pxl)\\
37 &  FUV\_ARTIFACT  & FUV artifact flag (logical OR near source)\\
38 & NUV\_ARTIFACT  & NUV  artifact flag (logical OR near source)\\
39 & FUV\_FLAGS  & FUV extraction flags  \\
40 & NUV\_FLAGS  & NUV extraction flags \\
41 & FUV\_FLUX  & FUV calibrated flux (micro Jy)\\
42 & FUV\_FLUXERR  &error on FUV calibrated flux (micro Jy) \\
43 & NUV\_FLUX  & NUV calibrated flux (micro Jy)\\
44 & NUV\_FLUXERR  & error on calibrated flux (micro Jy)\\
45 & FUV\_X\_IMAGE  & source position along X in FUV image \\
46 & FUV\_Y\_IMAGE  & source position along Y in FUV image\\
47 & NUV\_X\_IMAGE  & source position along X in NUV image\\
48 & NUV\_Y\_IMAGE  & source position along Y in NUV image\\
49 & FUV\_FWHM\_IMAGE  & FUV FWHM assuming a Gaussian core \\
50 & NUV\_FWHM\_IMAGE  & NUV FWHM assuming a Gaussian core \\
51 & FUV\_FWHM\_WORLD  & FUV FWHM assuming a Gaussian core (WORLD units)\\
52 & NUV\_FWHM\_WORLD  & NUV FWHM assuming a Gaussian core (WORLD units)\\
53 & NUV\_CLASS\_STAR  & S/G classifier output \\
54 & FUV\_CLASS\_STAR  &  S/G classifier output \\
55 & NUV\_ELLIPTICITY  & NUV 1.-B\_IMAGE/A\_IMAGE\\
56  & FUV\_ELLIPTICITY  & FUV 1.-B\_IMAGE/A\_IMAGE\\
57 & NUV\_THETA\_J2000 & NUV position angle (East of North) \\
58 &  NUV\_ERRTHETA\_J2000 & error on NUV position angle (East of North)\\
59 &  FUV\_THETA\_J2000  & FUV position angle (East of North)\\
60 & FUV\_ERRTHETA\_J2000  & error on FUV position angle (East of North)\\
61 & FUV\_NCAT\_FWHM\_IMAGE  & FUV FWHM\_IMAGE value from -fd-ncat.fits (pxl)\\
62 & FUV\_NCAT\_FLUX\_RADIUS\_3 &  FUV FLUX RADIUS \#3 (fd-ncat) (pxls) [0.80]\\
63 &  NUV\_KRON\_RADIUS & NUV Kron radius in units of A or B \\
64 &  NUV\_A\_WORLD  & NUV rms profile along major axis (WORLD units)\\
65 & NUV\_B\_WORLD  & NUV rms profile along minor axis (WORLD units) \\
66 & FUV\_KRON\_RADIUS & FUV  Kron radius in units of A or B \\
67 &  FUV\_A\_WORLD  & FUV rms profile along major axis (WORLD units)\\
68 & FUV\_B\_WORLD  & rms profile along minor axis (WORLD units) \\
69 & NUV\_WEIGHT  & NUV effective exposure (seconds)\\
70 & FUV\_WEIGHT  & FUV effective exposure (seconds)\\
71 & PROB  & probability of the FUV and NUV match \\
72 & SEP  & separation between FUV and NUV positions of the source (arcsec) \\
73 & NUV\_POSERR  & error on position of the NUV source (arcsec)\\
74 & FUV\_POSERR  & error on position of the FUV source (arcsec)\\
75 & IB\_POSERR  & interband position error (arcsec)\\
76 & NUV\_PPERR  & NUV Poisson position error (arcsec)\\
77 & FUV\_PPERR  & FUV Poisson position error (arcsec)\\
78 & CORV  & GUVcat tag - see \citet{bia17guvcat}\\
79 & GRANK  & GUVcat tag - see \citet{bia17guvcat}\\
80 & NGRANK  & GUVcat tag - see \citet{bia17guvcat}\\
81 & PRIMGID  & GUVcat tag - see \citet{bia17guvcat}\\
82 & GROUPGID  & GUVcat tag - see \citet{bia17guvcat}\\
83 & GRANKDIST  & GUVcat tag - see \citet{bia17guvcat}\\
84 & NGRANKDIST  & GUVcat tag - see \citet{bia17guvcat}\\
85 & PRIMGIDDIST  & GUVcat tag - see \citet{bia17guvcat}\\
86 & GROUPGIDDIST  & GUVcat tag - see \citet{bia17guvcat}\\
87 & GROUPGIDTOT  & GUVcat tag - see \citet{bia17guvcat}\\
88 & DIFFFUV  & GUVcat tag - see \citet{bia17guvcat}\\
89 & DIFFNUV  & GUVcat tag - see \citet{bia17guvcat}\\
90 & DIFFFUVDIST  & GUVcat tag - see \citet{bia17guvcat}\\
91 & DIFFNUVDIST  & GUVcat tag - see \citet{bia17guvcat}\\ 
92 & SEPAS  & separation between primary and secondary (arcsec) -see \citet{bia17guvcat}\\
93 & SEPASDIST  & separation between primary and secondary (arcsec), distance criterion -see \citet{bia17guvcat}\\
94 & INLARGEOBJ &  if the source is in the footprint of an extended object \\ 
 &  & INLARGEOBJ gives the object name (as in $GUVcat$) XX:name; where XX=GA (galaxy), \\ 
 & &  GC (globular cluster),  OC (open cluster), SC (other stellar clusters)\\
 & &  otherwise INLARGEOBJ is set to  ``N''   \\
95&  LARGEOBJSIZE &  size of the extended object: 1.25$\times$D25 for galaxies,\\ 
& &  2$\times$ radius for stellar clusters. LARGEOBJSIZE=0. if INLARGEOBJ = `N' \\
& & ( for sizes of the extended sources: see \citet{bia17guvcat}) \\ 
96$-$216 &   &  {\bf SDSS tags from GUVmatch} \tablenotemark{a}  \\
217 & DSTARCSEC & separation between GALEX and SDSS-match position (arcsec)\\ 
218 & DISTANCERANK & rank of multiple matches: =0 if this is the only SDSS match,\\
 & &  =1 if this is the closest (to GALEX source) of multiple SDSS matches \\
219 &  REVERSEDISTANCERANK & $>$0 if this SDSS source matches more than one GALEX source\\
220 &  MULTIPLEMATCHCOUNT & number of multiple SDSS matches to the GALEX source  \\
221 &  REVERSEMULTIPLEMATCHCOUNT & number of multiple GALEX sources matched by this SDSS source  \\
222 & GALEX\_JID & IAU-style source identifier  (from the GALEX coordinates)  \\
\multicolumn{3}{c}{\bf columns 223-231:  distilled information from the Simbad  match results}\\
223 & SIMBAD\_MATCH\_nearest & closest SIMBAD match (within 5\as)\tablenotemark{b}\\
224 & SIMBAD\_MAIN\_TYPE\_nearest & ``MAIN TYPE'' of  closest SIMBAD match (within 5\as)\tablenotemark{b}\\
225 & SIMBAD\_OTHER\_TYPES\_nearest & ``OTHER TYPES'' of the closest SIMBAD match within 5\as\tablenotemark{b} (stellar objects)\\
226 & SIMBAD\_SP\_TYPE\_nearest & spectral type of the closest SIMBAD match within 5\as\tablenotemark{b} \\
227 & SIMBAD\_ANGDIST\_nearest & distance (arcsec) of the closest SIMBAD match  (within 5\as)\tablenotemark{b} from the GALEX position\\
228 & N\_SIMBAD\_MATCHES\_3AS & PLACEHOLDER - not used - number of  SIMBAD matches within 3\as\\
228 & N\_SIMBAD\_MATCHES\_5AS & (NSIMBADMATCHES5AS)  number of  SIMBAD matches within 5\as\\ 
229 & N\_SIMBAD\_MATCHES\_10AS & (NSIMBADMATCHES10AS) number of  SIMBAD matches within 10\as\\
231 & N\_SIMBAD\_MATCHES\_30AS & PLACEHOLDER - not used - number of  SIMBAD matches within 30\as\\
\multicolumn{3}{c}{\bf columns 232-257:  main parameters of the GALEX ``best visit''}\\
232 & GALEX\_BESTVIS\_OBSDATE & date of GALEX best observation [asci]  (year-month-day, e.g. : 2005-11-05, \\
   & & or `` == `` if no visit found)\\ 
233 & GALEX\_BESTVIS\_NUVTIME & time of the NUV exposure of the GALEX best observation \\
  & & (hrs:min:sec, e.g., 08:23:51, or `` == `` if no visit found)\\ 
234 & GALEX\_BESTVIS\_FUVTIME & time of the FUV exposure of the GALEX best observation \\
 & & (hrs:min:sec, e.g., 08:23:51, or `` == `` if no visit found)\\  
235 & GALEX\_BESTVIS\_EPOCHDECIMAL & time of GALEX best observation, in year.decimal, \\
  & & for convenience of computing the difference with other epochs; \\ 
  & & e.g.,2003.7556152 (-888 for the 25 sources with no visit-level detection within 3\as)\\
  & &  Range: 2003.4315  to     2012.0968\\ 
236 & GALEX\_BESTVIS\_EPOCHNUVDECIMAL & time of GALEX best NUV observation, \\ 
  & & in year.decimal; range: 2003.4315$-$2012.0968\\
237 & GALEX\_BESTVIS\_EPOCHFUVDECIMAL & time of GALEX best FUV observation, \\
  & & in year.decimal; range: 2003.4315$-$2009.4059 among the 71148 visits with FUV detection; \\ 
  & & -888 for 216 visits with no FUV detection\\
238 & GALEX\_BESTVIS\_FUV\_WEIGHT & FUV effective exposure time of the GALEX best visit\tablenotemark{b} (seconds)\\ 
239 & GALEX\_BESTVIS\_NUV\_WEIGHT & NUV effective exposure time of the GALEX best visit\tablenotemark{b} (seconds)\\ 
240 & GALEX\_BESTVIS\_FOV\_RADIUS & distance of the source from the detector center (decimal degrees)\tablenotemark{b}\\ 
241 & GALEX\_BESTVIS\_RA & RA of the GALEX source measured in the best visit (decimal degrees)\tablenotemark{b}\\ 
242 & GALEX\_BESTVIS\_DEC & Dec of the GALEX source measured in the best visit (decimal degrees)\tablenotemark{b}\\ 
243 & GALEX\_BESTVIS\_FUVMAG & FUV mag of the GALEX source measured in the best visit (ABmag)\tablenotemark{b}\\ 
244 & GALEX\_BESTVIS\_FUVMAGERR & error of FUV mag of the GALEX source measured in the best visit (ABmag)\tablenotemark{b}\\ 
245 & GALEX\_BESTVIS\_NUVMAG &  NUV  mag of the GALEX source measured in the best visit (ABmag)\tablenotemark{b}\\ 
246 & GALEX\_BESTVIS\_NUVMAGERR & error of  NUV  mag of the GALEX source measured in the best visit (ABmag)\tablenotemark{b}\\ 
247 & GALEX\_NVIS\_FUV & number of FUV observations found [integer, 0 if no FUV visit found] \\ 
248 & GALEX\_NVIS\_NUV & number of NUV observations found [integer, 0 if no NUV visit found] \\ 
249 & GALEX\_MIN\_FUVMAG & minimum value of FUVMAG among all matched visits (excluding those where FUVMAG=-999) \\ 
250 & GALEX\_MIN\_FUVMAGERR & error of minimum  FUVMAG among all matched visits \\ 
251 & GALEX\_MAX\_FUVMAG & maximum value of FUVMAG among all matched visits\\ 
252 & GALEX\_MAX\_FUVMAGERR & error of maximum FUVMAG value  \\ 
253 & GALEX\_MIN\_NUVMAG & minimum value of NUVMAG among all matched visits\\ 
254 & GALEX\_MIN\_NUVMAGERR & error of minimum  NUVMAG among all matched visits \\ 
255 & GALEX\_MAX\_NUVMAG & maximum value of NUVMAG among all matched visits\\ 
256 & GALEX\_MAX\_NUVMAGERR & error of maximum NUVMAG value \\
257 & GALEX\_BESTVIS\_DISTARCSEC & separation between input coordinates and best-visit match of the GALEX source\\
\multicolumn{3}{c}{\bf tags selected from the Gaia DR3 match results (Gaia $source$ table) and multiple-matches tags}\\
258 & GAIA\_ID &  Gaia source\_ID of the closest Gaia DR3 match (-888 if no match)\\ 
259 &DISTARCMIN & distance betwee the Gaia counterpart and the GALEX source (arcmin)\\ 
260 &GAIA\_RA & RA of the Gaia counterpart (decimal degrees) \\ 
261 &GAIA\_DEC & DEC of the Gaia counterpart (decimal degrees) \\ 
262 &MMRANK\_GAIA & multiple-match rank (Section \ref{s_gaia})\\ 
263 &RMMRANK\_GAIA & reverse multiple-match rank (Section \ref{s_gaia})\\ 
264 &NMMRANK\_GAIA & number of Gaia DR3 matches within the match radius\\ 
265 & RNMMRANK\_GAIA & reverse number of matches (of this Gaia source to other GALEX sources)\\ 
266 & PARALLAX &  Gaia DR3 parallax (mas)\\ 
267 & PARALLAX\_ERROR &  error on Gaia DR3 parallax (mas)\\ 
268 & PARALLAX\_OVER\_ERROR & parallax/error from Gaia DR3 \\
269 &  PM & proper motion from Gaia DR3 (mas) \\ 
270 & PMRA & RA component of proper motion from Gaia DR3 (mas)\\ 
271 & PMRA\_ERROR &  error on RA component of proper motion from Gaia DR3 (mas) \\ 
272 & PMDEC & Dec component of proper motion from Gaia DR3 (mas)\\ 
273 & PMDEC\_ERROR & error on Dec component of proper motion from Gaia DR3 (mas)\\ 
274 & ASTROMETRIC\_N\_GOOD\_OBS\_AL & Number of Gaia observations for the solution\\ 
275 & RUWE & from Gaia DR3  {\it source} table  \\ 
276 & ASTROMETRIC\_EXCESS\_NOISE & from Gaia DR3 {\it source} table  \\ 
277 & ASTROMETRIC\_EXCESS\_NOISE\_SIG & from Gaia DR3 {\it source} table  \\ 
278 & DUPLICATED\_SOURCE PHOT\_G\_MEAN\_MAG& from Gaia DR3 {\it source} table \\ 
279 & PHOT\_G\_MEAN\_FLUX\_OVER\_ERROR & from Gaia DR3 {\it source} table \\ 
280 & PHOT\_BP\_MEAN\_MAG & from Gaia DR3 {\it source} table \\ 
281 & PHOT\_BP\_MEAN\_FLUX\_OVER\_ERROR & from Gaia DR3 {\it source} table \\ 
282 & PHOT\_RP\_MEAN\_MAG & from Gaia DR3 {\it source} table \\ 
283 & PHOT\_RP\_MEAN\_FLUX\_OVER\_ERROR & from Gaia DR3 {\it source} table \\ 
284 & RADIAL\_VELOCITY& from Gaia DR3 {\it source} table  \\ 
285 & RADIAL\_VELOCITY\_ERROR & from Gaia DR3 {\it source} table \\ 
\multicolumn{3}{c}{\bf  tags selected from Gaia DR3 table {\it vary\_summary} } \\
286 & VARI\_SUMMARY\_GAIA\_ID &  from Gaia DR3 table {\it vari\_summary} \\ 
287 & PHOT\_VARIABLE\_FLAG & from Gaia DR3 table {\it vari\_summary} \\ 
288 & NON\_SINGLE\_STAR & from Gaia DR3 table {\it vari\_summary} \\ 
289 & IN\_QSO\_CANDIDATES & from Gaia DR3 table {\it vari\_summary} \\ 
290 & IN\_GALAXY\_CANDIDATES & from Gaia DR3 table {\it vari\_summary} \\ 
291 & MIN\_MAG\_G\_FOV &from Gaia DR3 table {\it vari\_summary} \\
292 & MAX\_MAG\_G\_FOV & from Gaia DR3 table {\it vari\_summary} \\ 
293 & MEAN\_MAG\_G\_FOV & from Gaia DR3 table {\it vari\_summary} \\ 
294 & MEDIAN\_MAG\_G\_FOV & from Gaia DR3 table {\it vari\_summary} \\ 
295 & RANGE\_MAG\_G\_FOV & from Gaia DR3 table {\it vari\_summary} \\ 
296 & TRIMMED\_RANGE\_MAG\_G\_FOV & from Gaia DR3 table {\it vari\_summary} \\ 
297 & STD\_DEV\_MAG\_G\_FOV & from Gaia DR3 table {\it vari\_summary} \\ 
298 & SKEWNESS\_MAG\_G\_FOV & from Gaia DR3 table {\it vari\_summary} \\ 
299 & KURTOSIS\_MAG\_G\_FOV& from Gaia DR3 table {\it vari\_summary} \\ 
300 & IN\_VARI\_CLASSIFICATION\_RESULT & values can be False, True, or -888 if no match exists \\ 
301 & IN\_VARI\_RRLYRAE & values can be False, True, or -888 if no match exists\\ 
302 & IN\_VARI\_CEPHEID & values can be False, True, or -888 if no match exists\\ 
303 & IN\_VARI\_PLANETARY\_TRANSIT & values can be False, True, or -888 if no match exists\\ 
304 & IN\_VARI\_SHORT\_TIMESCALE & values can be False, True, or -888 if no match exists\\ 
305 & IN\_VARI\_LONG\_PERIOD\_VARIABLE & values can be False, True, or -888 if no match exists\\ 
306 & IN\_VARI\_ECLIPSING\_BINARY & values can be False, True, or -888 if no match exists\\ 
307 & IN\_VARI\_ROTATION\_MODULATION & values can be False, True, or -888 if no match exists\\ 
308 & IN\_VARI\_MS\_OSCILLATOR & values can be False, True, or -888 if no match exists\\ 
309 & IN\_VARI\_AGN & values can be False, True, or -888 if no match exists\\ 
310 & IN\_VARI\_MICROLENSING & values can be False, True, or -888 if no match exists\\ 
311 & IN\_VARI\_COMPACT\_COMPANION & values can be False, True, or -888 if no match exists\\
\multicolumn{3}{c}{\bf  distance and range computed from Gaia parallax and parallax\_error} \\
312 & DISTANCE\_GAIA\_PC & distance (in pc) from parallax (value = -888.0 if no match exists)\\ 
313 & DISTANCE\_GAIA\_ERRPLUS & maximum distance (in pc) applying parallax error \\ 
314 & DISTANCE\_GAIA\_ERRMINUS & minimum distance (in pc) applying parallax error\\ 
\enddata
\tablenotetext{*}{These columns are available for all sources, even those that are eliminated from the analysis}
\tablenotetext{a}{{\tiny 
SDSS\_ID NCHILD SDSS\_RA SDSS\_RA\_ERROR SDSS\_DEC SDSS\_DEC\_ERROR SDSS\_TYPE PSFMAG\_U PSFMAG\_G PSFMAG\_R PSFMAG\_I PSFMAG\_Z PSFMAGERR\_U
PSFMAGERR\_G PSFMAGERR\_R PSFMAGERR\_I PSFMAGERR\_Z EXPMAG\_U EXPMAG\_G EXPMAG\_R EXPMAG\_I EXPMAG\_Z EXPMAGERR\_U EXPMAGERR\_G EXPMAGERR\_R EXPMAGERR\_I EXPMAGERR\_Z EXPAB\_U EXPAB\_G EXPAB\_R EXPAB\_I
EXPAB\_Z EXPABERR\_U EXPABERR\_G EXPABERR\_R EXPABERR\_I EXPABERR\_Z DEVMAG\_U DEVMAG\_G DEVMAG\_R DEVMAG\_I DEVMAG\_Z DEVMAGERR\_U DEVMAGERR\_G DEVMAGERR\_R DEVMAGERR\_I DEVMAGERR\_Z DEVAB\_U DEVAB\_G
DEVAB\_R DEVAB\_I DEVAB\_Z DEVABERR\_U DEVABERR\_G DEVABERR\_R DEVABERR\_I DEVABERR\_Z PETROMAG\_U PETROMAG\_G PETROMAG\_R PETROMAG\_I PETROMAG\_Z PETROMAGERR\_U PETROMAGERR\_G PETROMAGERR\_R
PETROMAGERR\_I PETROMAGERR\_Z FLAGS1 FLAGS2 FLAGS\_U FLAGS\_G FLAGS\_R FLAGS\_I FLAGS\_Z EDGE U\_SAT G\_SAT R\_SAT I\_SAT Z\_SAT U\_CR G\_CR R\_CR I\_CR Z\_CR PROBPSF\_U PROBPSF\_G PROBPSF\_R PROBPSF\_I
PROBPSF\_Z PSFFWHM\_U PSFFWHM\_G PSFFWHM\_R PSFFWHM\_I PSFFWHM\_Z SPECOBJ\_ID PLATE MJD FIBER\_ID REDSHIFT REDSHIFTERR SPECTYPE CLASS SUBCLASS CLASS\_PERSON ELODIESPTYPE B\_V\_COLOR TEFF LOGG
METALLICITY ELODIE\_REDSHIFT ELODIE\_REDSHIFTERR PROPERMOTION USNO\_RED1 USNO\_RED2 USNO\_BLUE1 USNO\_BLUE2 RUN RERUN CAMCOL FIELD}} 
\tablenotetext{c}{value is =-888 if no match is found}
\end{deluxetable} 
\clearpage 

\begin{deluxetable}{lll}
\tabletypesize{\tiny}
\tablecaption{Description of Columns of Gaia matched catalog\tablenotemark{*}  \label{t_gaiatags} }
\tablehead{
\colhead{Col} & \colhead{Column Name} & \colhead{Description} }
\startdata
\tabletypesize{\tiny}
1-264 &  & columns from Table \ref{t_tagspoint}\\
265-440 & & all tags from Gaia DR3 $source$ and $vari\_summary$ tables for each match\\
441  & GAIA\_ID &  SOURCE\_ID of the Gaia match  in Gaia DR3 $source$ table\\
442  & DISTARCMIN  & separation (arcmin) between the GALEX source position and the Gaia match\\
443  & GAIA\_RA & RA (deg ) of Gaia match  \\
444  & GAIA\_DEC &  DEC (deg) of Gaia match  \\
445  & VARI\_SUMMARY\_GAIA\_ID & ID of the Gaia match in the DR3 table $vari\_summary$\\
446  & VARI\_SUMMARY\_SOLUTION\_ID &  SOLUTION\_ID from linking the DR3 table $vari\_summary$\\ 
447  & MMRANK\_GAIA  & =0 if this Gaia match is the only one, \\
     &               & =1 if this Gaia match is the closest of more than one match \\
     &               & $>$1 in order of distance, for additional matches\\
     &               & =-99 if no Gaia match was found within the match radius\\
448  & RMMRANK\_GAIA & reverse match rank (tracks if a Gaia source matches more than one input source) \\ 
449  & NMMRANK\_GAIA  & number of Gaia matches to the GALEX source\\
450  & RNMMRANK\_GAIA & number of reverse matches count (how many GALEX sources does this Gaia source match)\\
\enddata
\end{deluxetable}

{}
  
\acknowledgments

LB acknowledges support from NASA ADAP grants 80NSSC19K0527 and NNX17AF35G.  I am grateful to the referee whose comments prompted  useful clarifications of the manuscript, to Bernie Shiao for helpful discussions about issues in the GALEX database, to Kareem El-Badry for insightful suggestions concerning Gaia data and the science results, and to Stavros Akras and Anastasios Karagiannis for  beta-testing the catalog for possible applications. 

This work includes results from the European Space Agency (ESA) space mission Gaia, Data Release 3. Gaia data are being processed by the Gaia Data Processing and Analysis Consortium (DPAC). Funding for the DPAC is provided by national institutions, in particular the institutions participating in the Gaia MultiLateral Agreement (MLA). The Gaia mission website is https://www.cosmos.esa.int/gaia.   Gaia data were accessed through the MAST interface.

{\it Facilities:} \facility{GALEX},  \facility{MAST} 

\textbf{ORCID iDs}
Luciana Bianchi https://orcid.org/0000-0001-7746-5461

\end{document}